\documentclass[nopreprintline,final,5p,times,twocolumn]{elsarticle}

\usepackage{amssymb}

\usepackage[utf8]{inputenc}
\usepackage{subcaption}
\usepackage{amsmath}
\usepackage{mathtools}

\usepackage{xspace} 
\usepackage{mathrsfs} 
\usepackage[inline]{enumitem} 
\usepackage{booktabs}
\usepackage{nicefrac}
\usepackage{hyperref}
\usepackage{kantlipsum,cuted}

\usepackage{siunitx}
\usepackage{color}

\newcommand{\smmu}{\ensuremath{\mu_{\mathrm{s}}}\xspace}

\newcommand{\Tc}{\ensuremath{T_{\mathrm{c}}}\xspace}
\newcommand{\kB}{\ensuremath{k_{\mathrm{B}}}\xspace}

\newcommand{\Mags}{\ensuremath{M_{\mathrm{s}}}\xspace}

\newcommand{\Tamb}{\ensuremath{T_{\mathrm{a}}}\xspace}

\newcommand{\Tpeak}{\ensuremath{T_{\mathrm{peak}}}\xspace}
\newcommand{\tpulse}{\ensuremath{t_{\mathrm{pulse}}}\xspace}

\newcommand{\zetaad}{\ensuremath{\mathbf{\zeta}_{ad}}\xspace}
\newcommand{\zetaperp}{\ensuremath{\mathbf{\zeta}_{\bot}}\xspace}
\newcommand{\alphapara}{\ensuremath{\alpha_{\parallel}}\xspace}
\newcommand{\alphaperp}{\ensuremath{\alpha_{\bot}}\xspace}
\newcommand{\mvec}{\ensuremath{\mathbf{m}}\xspace}
\newcommand{\Hintragrainvec}{\ensuremath{\mathbf{H}_{\mathrm{intragrain}}}\xspace}

\newcommand{\Hexchangevec}{\ensuremath{\mathbf{H} }_{\mathrm{exch}}\xspace}
\newcommand{\Chipara}{\ensuremath{\Tilde{\chi}_{\parallel}}\xspace}
\newcommand{\Chilong}{\ensuremath{\Tilde{\chi}_{l}}\xspace}
\newcommand{\Chiperp}{\ensuremath{\Tilde{\chi}_{\bot}}\xspace}
\newcommand{\Chiparaperp}{\ensuremath{\Tilde{\chi}_{\parallel,\perp}}\xspace}
\newcommand{\Hsat}{\ensuremath{H_{\mathrm{sat}}}\xspace}
\newcommand{\Heffvec}{\ensuremath{\mathbf{H}_{\mathrm{eff}}}\xspace}
\newcommand{\Hdmagvec}{\ensuremath{ \mathbf{H} }_{\mathrm{dmg}}\xspace}
\newcommand{\Hthvec}{\ensuremath{\mathbf{H} }_{\mathrm{th}}\xspace}
\newcommand{\Happvec}{\ensuremath{\mathbf{H} }_{\mathrm{app}}\xspace}
\newcommand{\Hanivec}{\ensuremath{\mathbf{H} }_{\mathrm{ani}}\xspace}

\journal{}

\begin{document}

\begin{frontmatter}



\title{Models of Advance Recording Systems: A Multi-timescale Micromagnetic code for granular thin film magnetic recording systems}
\author[inst1]{S. E. Rannala\corref{cor1}}
\ead{ser527@york.ac.uk}
\cortext[cor1]{Corresponding authors}
\author[inst1,inst2]{A. Meo}
\author[inst1,inst3]{S. Ruta}
\author[inst2]{W. Pantasri}
\author[inst1]{R. W. Chantrell}
\author[inst2]{P. Chureemart}
\author[inst2]{J. Chureemart\corref{cor1}}
\ead{jessada.c@msu.ac.th}
\affiliation[inst1]{organization={Department of Physics},
            addressline={University of York},
            city={York},
            postcode={YO10 5BP},
            country={U.K.}}
\affiliation[inst2]{organization={Department of Physics},
            addressline={Mahasarakham University},
            city={Mahasarakham},
            postcode={44150},
            country={Thailand}}
\affiliation[inst3]{organization={College of Business, Technology and Engineering},
            addressline={Sheffield Hallam University},
            city={Sheffield},
            country={U.K.}}

\begin{abstract}
Micromagnetic modelling provides the ability to simulate large magnetic systems accurately without the computational cost limitation imposed by atomistic modelling. Through micromagnetic modelling it is possible to simulate systems consisting of thousands of grains over a time range of nanoseconds to years, depending upon the solver used. Here we present the creation and release of an open-source multi-timescale micromagnetic code combining three key solvers: Landau-Lifshitz-Gilbert; Landau-Lifshitz-Bloch; Kinetic Monte Carlo. This code, called MARS (Models of Advanced Recording Systems), is capable of accurately simulating the magnetisation dynamics in large and structurally complex single- and multi-layered granular systems. The short timescale simulations are achieved for systems far from and close to the Curie point via the implemented Landau-Lifshitz-Gilbert and Landau-Lifshitz-Bloch solvers respectively. This enables read/write simulations for general perpendicular magnetic recording and also state of the art heat assisted magnetic recording (HAMR). The long timescale behaviour is simulated via the Kinetic Monte Carlo solver, enabling investigations into signal-to-noise ratio and data longevity. The combination of these solvers opens up the possibility of multi-timescale simulations within a single software package. For example the entire HAMR process from initial data writing and data read back to long term data storage is possible via a single simulation using MARS. The use of atomistic parameterisation for the material input of MARS enables highly accurate material descriptions which provide a bridge between atomistic simulation and real world experimentation. Thus MARS is capable of performing simulations for all aspects of recording media research and development. This ranges from material characterisation and optimisation to system design and implementation. The object orientated nature of MARS is structured to facilitate quick and simple development and easy implementation of user defined custom simulation types which can utilise either timescale or a combination of both timescales.
\end{abstract}


\end{frontmatter}

{\bf PROGRAM SUMMARY} \\
\begin{small}
\noindent
{\em Program Title:} MARS                                          \\
{\em Developer's repository link:} \url{https://bitbucket.org/EwanRannala/mars/} \\
{\em Licensing provisions:} MIT  \\
{\em Programming language:} C++                                 \\
{\em Supplementary material:} MARS testing methodology\\
{\em Nature of problem:} A combined model that enables the complete modelling of magnetic recording processes at elevated temperatures covering all time scales from writing (nanoseconds) up to long term data storage (years). The model must also accurately describe the granular nature of the recording media as grain sizes are reduced to a few nanometres. 
\\
{\em Solution method:} Short timescale behaviours are captured via the Landau-Lifshitz-Gilbert and Landau-Lifshitz-Bloch solvers for low and high temperature systems respectively. The long time scale behaviours is captured via a kinetic Monte Carlo solver. To enable complex models which account for mixed timescale behaviours the solvers are implemented as a single class structure which allows for dynamic solver selection. The granular structure is generated via a Laguerre-Voronoi tessellation with a custom implemented packing algorithm to produce highly realistic grain size distributions. Complex thermal dependencies of materials can be incorporated via atomistic parameterisation forming a multi-timescale model of the material.
\end{small}

\section{Introduction}
Magnetic recording using hard disk drives remains the dominant technology for cloud-based information storage. As data centres consume sufficient energy to represent a significant contribution to global warming there is an imperative to improve energy efficiency and minimise the required number of data centres by means of increasing storage density. In order to achieve the properties required for magnetic information storage, the storage medium must be granular in nature. The essential required property for the storage of binary information is the presence of a large magnetic anisotropy ($K$): the material property which provides an energy barrier to switching of the magnetisation and thereby creates a two-state magnetic system. 
The increase of areal density generally proceeds by a scaling of properties, particularly a reduction of the grain size to ensure adequate signal to noise ratio (SNR) given the reduction in bit size. This necessitates an increase in the value of $K$ to ensure thermal stability, which makes the writing of individual bits more difficult due to an increase in write field: the magnetic `trilemma' \cite{Richter2007}. Current technology (perpendicular magnetic recording) is already running into write-field limitations and a step change of technology is required for future products. Based on the ASRC (Advanced Storage Research Consortium) road map, there are two future technologies: 1) heat assisted magnetic recording (HAMR) and 2) Bit patterned magnetic recording (BPMR), which combined can lead to Heated-Dot magnetic recording \cite{IDEMA}. To facilitate the development and optimisation of the present-day and future magnetic recording technologies advanced models with greater levels of complexity are required. These models must capture the dynamics at both short and long timescales over a large number of bits/grains whilst accurately describing the variation of magnetic fields, temperatures and temperature dependent parameters.

HAMR is currently the most promising new technology to provide recording densities significantly greater than those available via current standard perpendicular magnetic recording \cite{SEAGATE}. HAMR provides increased areal densities through the utilisation of high coercivity materials, potentially enabling up to $4Tb/in\textsuperscript{2}$ \cite{Weller2014}.
The initial difficulty with using high coercivity materials is the requirement of increased writing head field gradients. The solution to this is to temporarily reduce the  coercivity of the medium, thus enabling writing with reduced field strengths. This process is achieved by applying a laser pulse to heat the material and cause a reduction in the material anisotropy. 
The physics of HAMR, involving heating up to or beyond the magnetic ordering (Curie) temperature $T_c$ remains challenging and involves models with a thermodynamic basis beyond the scope of those used in the typical micromagnetic approach.

Atomistic models provide this level of detail and have been used to provide temperature dependent magnetic properties and reversal mechanisms in recording media \cite{Evans2014_field_cooling,Vogler2016_AD_optimise,Vogler2016_noise, chureemart2017hybrid,ababei2019anomalous,muthsam2019improving, Ellis2017,strungaru2020model,meo2020magnetisation, waters2019resolving}. Although atomistic simulations can provide exceptional detail of the underlying physical processes which govern their macroscopic properties these simulations carry a significant computational cost. This cost limits atomistic simulations of recording media to a lengthscale of a few grains and a timescale of nanoseconds. 
This significant computational cost limits the investigation of statistical variation in particle properties and inter-particle interaction or temperature/field profiles over a track of recording bits. 
However, to be able to design, test and optimise any present-day and future magnetic recording technology, it is vital to capture the recording of thousands of grains from the sub-nanosecond timescale to a data storage timescale (5-10 years). 

Furthermore, experimental characterisations and tests are performed at the nanosecond timescale for FMR and millisecond timescale for standard measurements such as hysteresis loops, thermal decay, First Order Reversal Curves (FORC) and thermoremanence \cite{Richardson2018,Tc2014}. Over long timescales, thermally activated transitions over the energy barriers can lead to loss of recorded information. Transition times are governed by the Arrhenius-N\'{e}el law \cite{NEEL}, which gives a characteristic time $\tau^{-1}=f_0\exp(KV/kT)$, where $V$ is the grain volume and $K$ is the magnetic anisotropy constant. The pre-exponential factor is dependent on the value of $K$ and is in the region of \SI{}{GHz} to \SI{}{THz}. Ensuring thermal stability of the written information for 5-10 years requires large energy barriers ($KV/kT>80$) and good SNR. Typically short and long timescale investigations are performed separately, mainly due to the described fundamental difference between the simulation methodology. Nonetheless, there are numerous cases where both timescales are of interest, the simplest example of this is the effect of the writing process on nearby bits (nanosecond timescale) and data longevity ( timescale of years). 

\begin{figure}
    \centering
    \includegraphics[width=\linewidth]{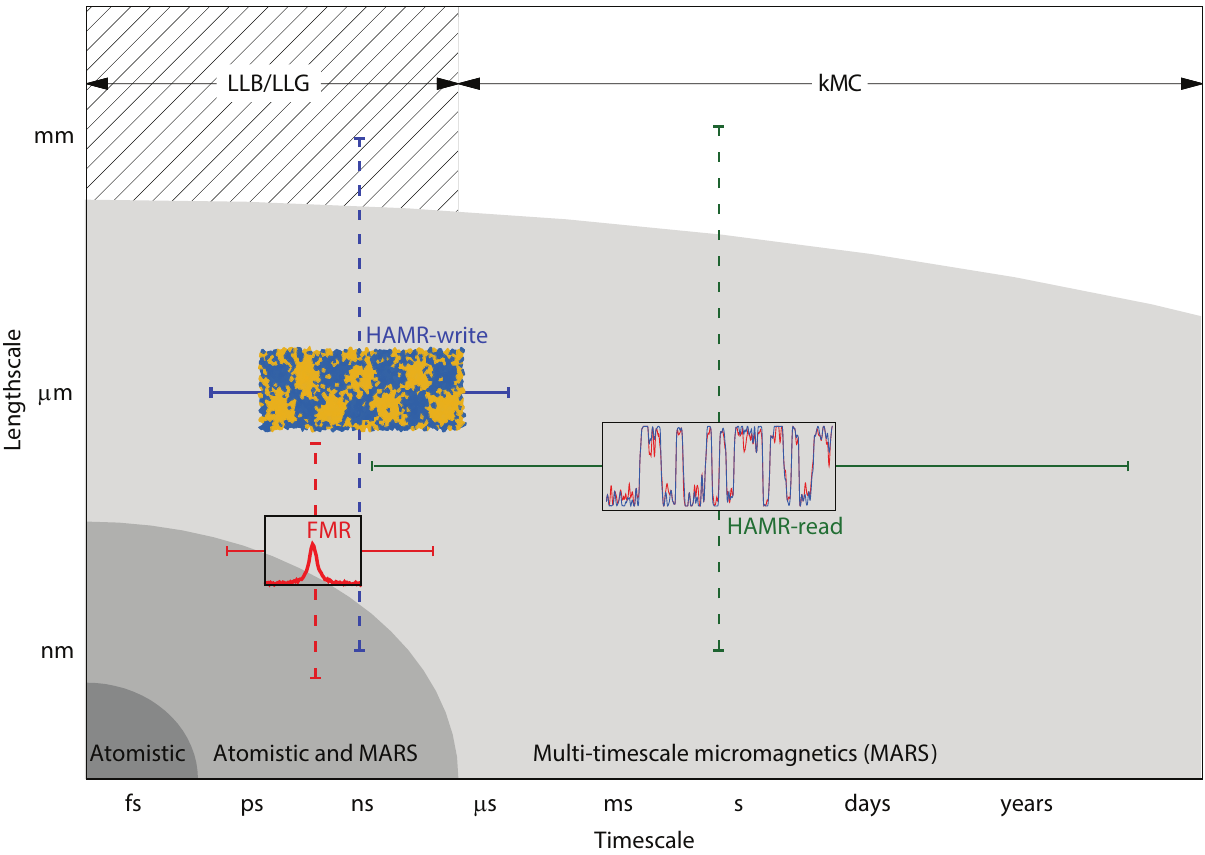}
    \caption{Illustration of the time/length scales required by some common simulations.}
    \label{fig:Scales}
\end{figure}

Here we present the developed multi-timescale micromagnetic code, MARS (Models of Advanced Recording Systems), which has the functionality of utilising various solvers to best accommodate the required simulation time frames.
MARS includes both short and long timescale solvers to make such investigations more simple and to open up the possibility to more easily access the effect of dynamic processes on the long timescale behaviour. MARS includes a stochastic-Landau-Lifshitz-Gilbert (sLLG) equation solver along with a stochastic-Landau-Lifshitz-Bloch solver, specifically the sLLB-II (which is shown to have greater accuracy at the Curie temperature \cite{sLLB-II}), to simulate short timescale dynamics of the magnetisation. A kinetic Monte-Carlo (kMC) solver is also included in order to simulate the long timescale behaviour. Fig. \ref{fig:Scales} illustrates the benefits of the multi-timescale micromagnetic approach over computationally expensive atomistic simulations. The addition of the kMC solver enables the simulation of timescales far beyond the standards set using dynamical micromagnetic solvers alone. 

MARS has been designed to be used alongside atomistic simulations by utilising parameterisation obtained via atomistic simulation to describe material properties, this allows for highly accurate descriptions of simulated materials. There is no limit on the number of materials MARS can utilise in a single simulation. This enables multi-layered systems such as those used for exchange-coupled-composites to be simulated. 
Details of granular system generation and the numerical solvers are provided in section \ref{sec:framework}.
MARS includes a set of commonly used simulation types for easy use, these include: HAMR writing, time evolution and data read back, thermoremanence, FORC and FMR. A selection of results obtained via these simulations are provided and discussed in section \ref{sec:Simulations}.

\section{MARS: multiscale framework}
\label{sec:framework}

\subsection{Granular model}

The micromagnetic approach consists of treating magnetic grains as macrospins with associated magnetisation, \mvec. To ensure accurate modelling of these systems it is crucial to generate accurate granular structures. The typical method for generating granular structures is via Voronoi tessellation \cite{Hai2014,Zhu2013,Eason2014,Victoria2013}. The classical Voronoi algorithm starts with the creation of seed points throughout the system, cell walls are then created such that they lie halfway between two seed points. While the general process is the same there exist various methods to determine the initial locations of the seed points \cite{Zhu20012,GRENESTEDT1998}.

A major drawback with the classical Voronoi construction is evident when a distribution of grain sizes is required. When there is a local increase in seed density the construction can generate unrealistically angular cells. This occurs due to the only constraint on cell construction being the requirement that the cell wall must be equidistant between seed points.

\begin{figure}
    \centering
    \includegraphics[width=\linewidth]{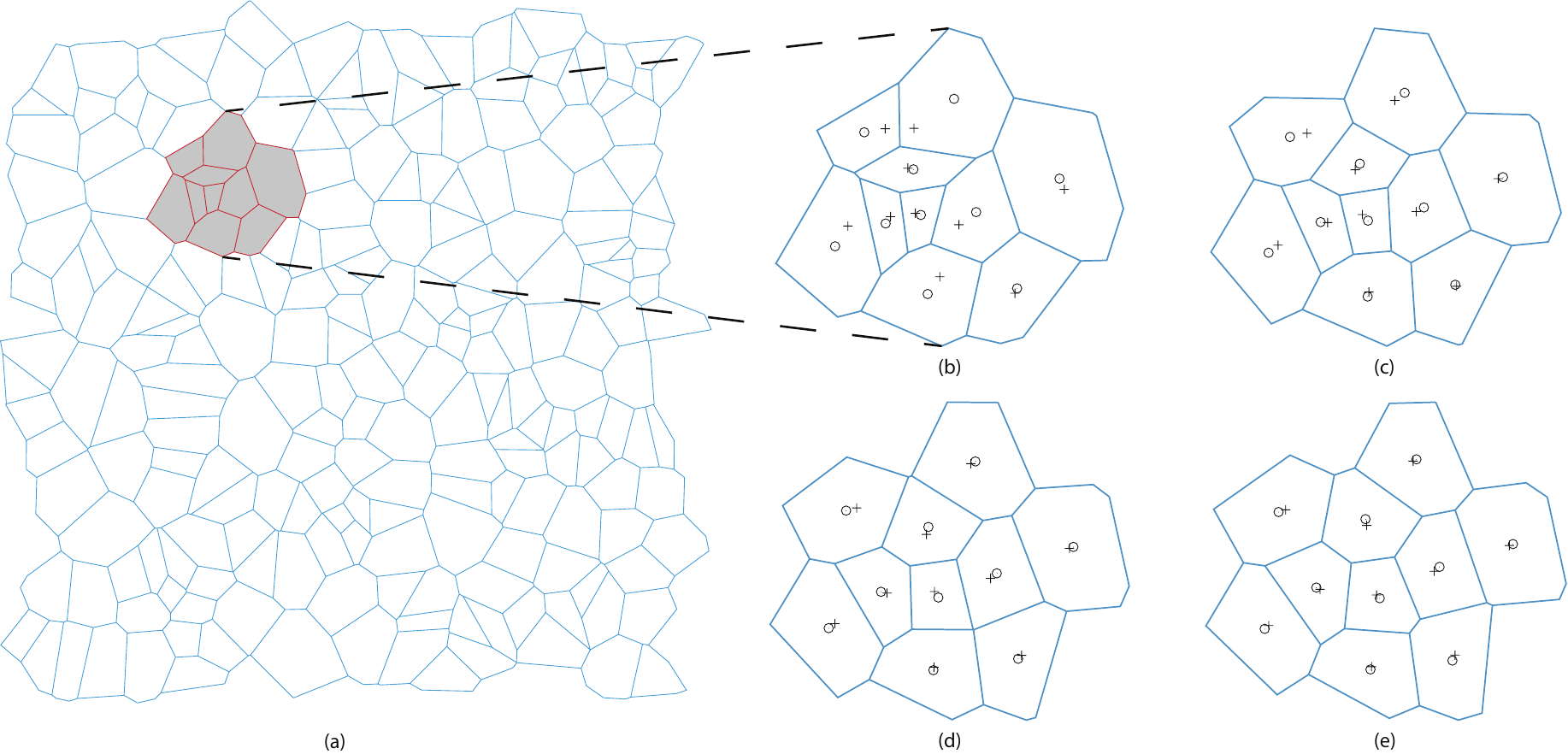}
    \caption{Example of the granular structure obtained via the centroidal Voronoi tessellation (a). A subsection has been chosen to illustrate Lloyd's relaxation algorithm after: zero (b), one (c), two (d) and three (e) iterations. The seed points are indicated by the crosses with the centroids represented by the circles. As more iterations are performed the angular nature of the grains is reduced.}
    \label{fig:Lloyd}
\end{figure}

To overcome this drawback, one can utilise centroidal Voronoi tessellation, this modified Voronoi process has been shown to be most effective when combined with Lloyd's algorithm \cite{Lloyd,Liu2009}. The process involves iterative relaxation of the constructed granular system by replacing the initial seed points with the centroids (typically known as the centre of mass) of the generated cells until convergence is achieved. Fig. \ref{fig:Lloyd} shows the system generated via MARS using centroidal Voronoi tessellation followed by the changes produced by applying Lloyd's algorithm over three iterations. 

\begin{figure}[ht]
    \centering
    \includegraphics[width=\linewidth]{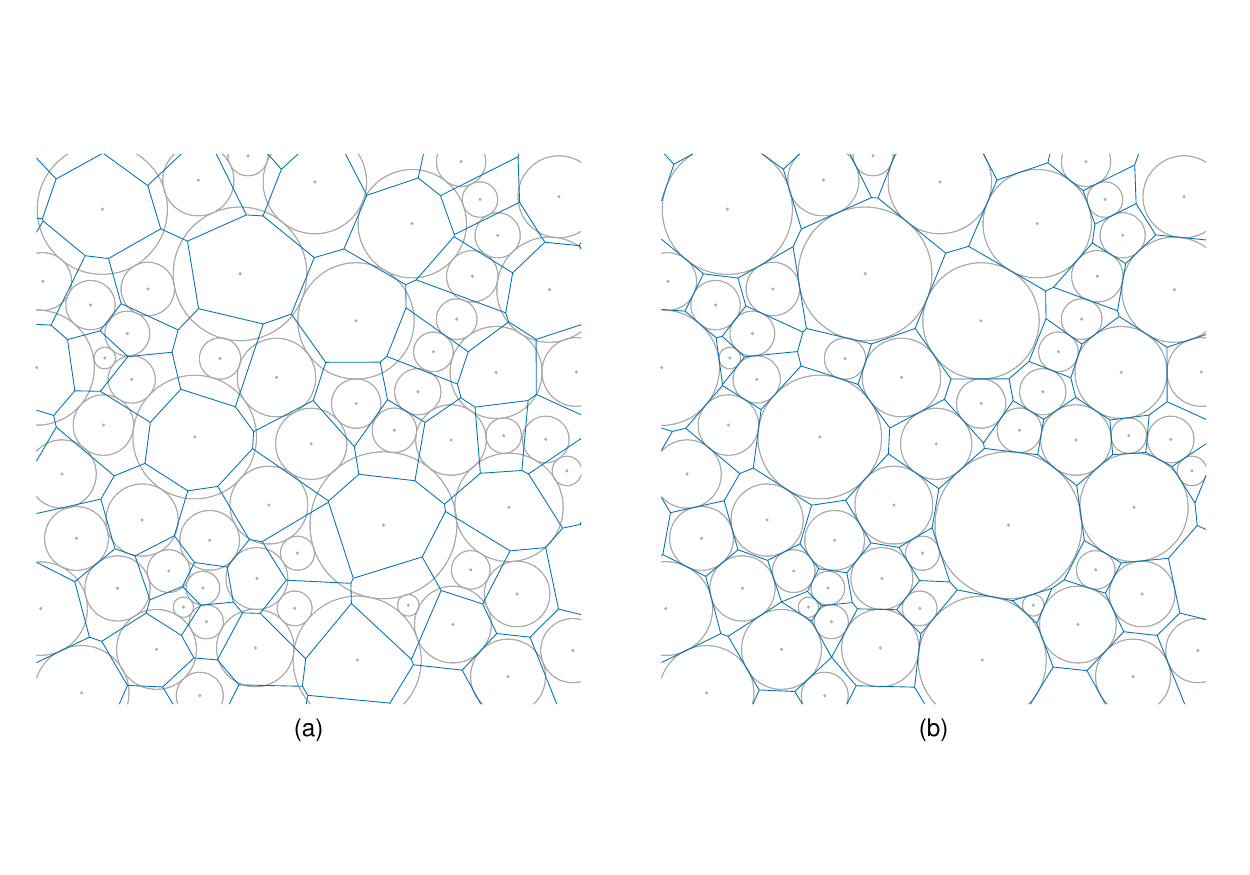}
    \caption{Granular system generated from an single arrangement of hard discs via the centroidal Voronoi tessellation (a) and the Laguerre-Voronoi tessellation (b).}
    \label{fig:V_v_LV_bub}
\end{figure}

\begin{figure}[ht]
    \centering
    \includegraphics[width=\linewidth]{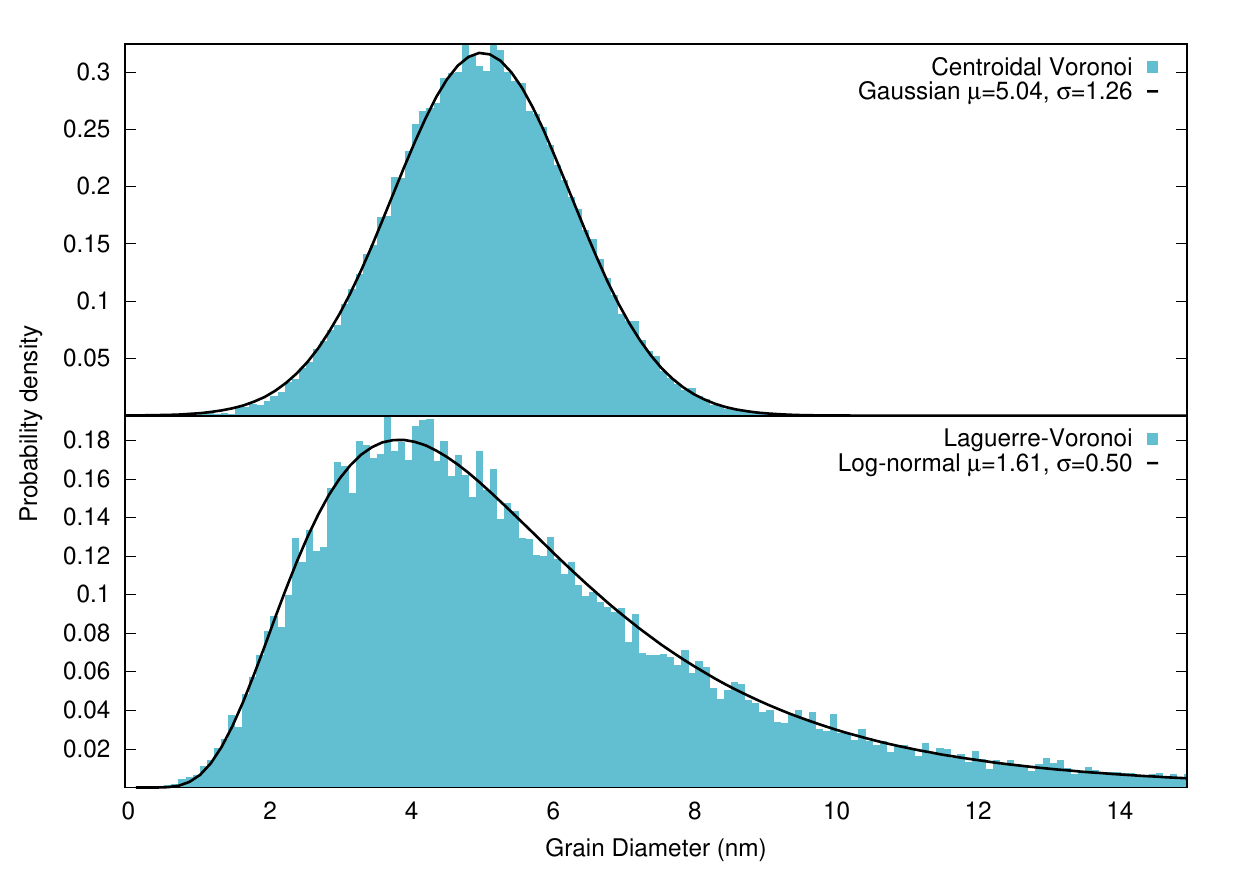}
    \caption{Grain diameter distributions for centroidal Voronoi tessellation (top) and Laguerre-Voronoi tessellation (bottom). The input distribution was lognormal with $\mu=1.65$ and $\sigma=0.55$. The seed based Voronoi is unable to provide the desired distribution, instead providing a Gaussian. The Laguerre-Voronoi was able to produce the desired distribution type with only a small change of parameters, which is to be expected due to the random nature of the packing process.}
    \label{fig:Dist_comp}
\end{figure}

A key limitation to all seed-based construction techniques is that they produce Gaussian cell size distributions \cite{FAZEKAS2002,Garvois}. In reality the grain size distribution has been shown to be described best by lognormal or Gamma distributions \cite{Su-Peng2009,Weller2014,Ganping2006}. To enable the construction of systems following these distributions a method for performing a Voronoi construction using `hard discs' instead of seeds is available, called the Laguerre-Voronoi method. The main challenge to the Laguerre-Voronoi method is the initial packing of the randomly sized hard discs. There are numerous methods for packing these hard discs, for MARS a custom ``Drop and Roll'' method has been implemented. This method provides a high level of contact between neighbouring discs resulting in a greater packing fraction while being more computationally efficient \cite{ALSAYEDNOOR2016}. Using this algorithm MARS is capable of generating packing fractions of at least 80\% for randomly distributed disc sizes. Once the system has been packed the cells are generated using the robust open source VORO++ package developed by Rycroft \cite{VORO}. Fig. \ref{fig:V_v_LV_bub} shows the difference between the structures generated via the centroidal Voronoi tessellation and the Laguerre-Voronoi tessellation, the corresponding grain size distributions are shown in Fig. \ref{fig:Dist_comp}. The implementation of the Laguerre-Voronoi tessellation method within MARS enables the generation of realistic systems with a high level of control over the grain size distributions.

\begin{figure}[ht]
    \centering
    \includegraphics[width=\linewidth]{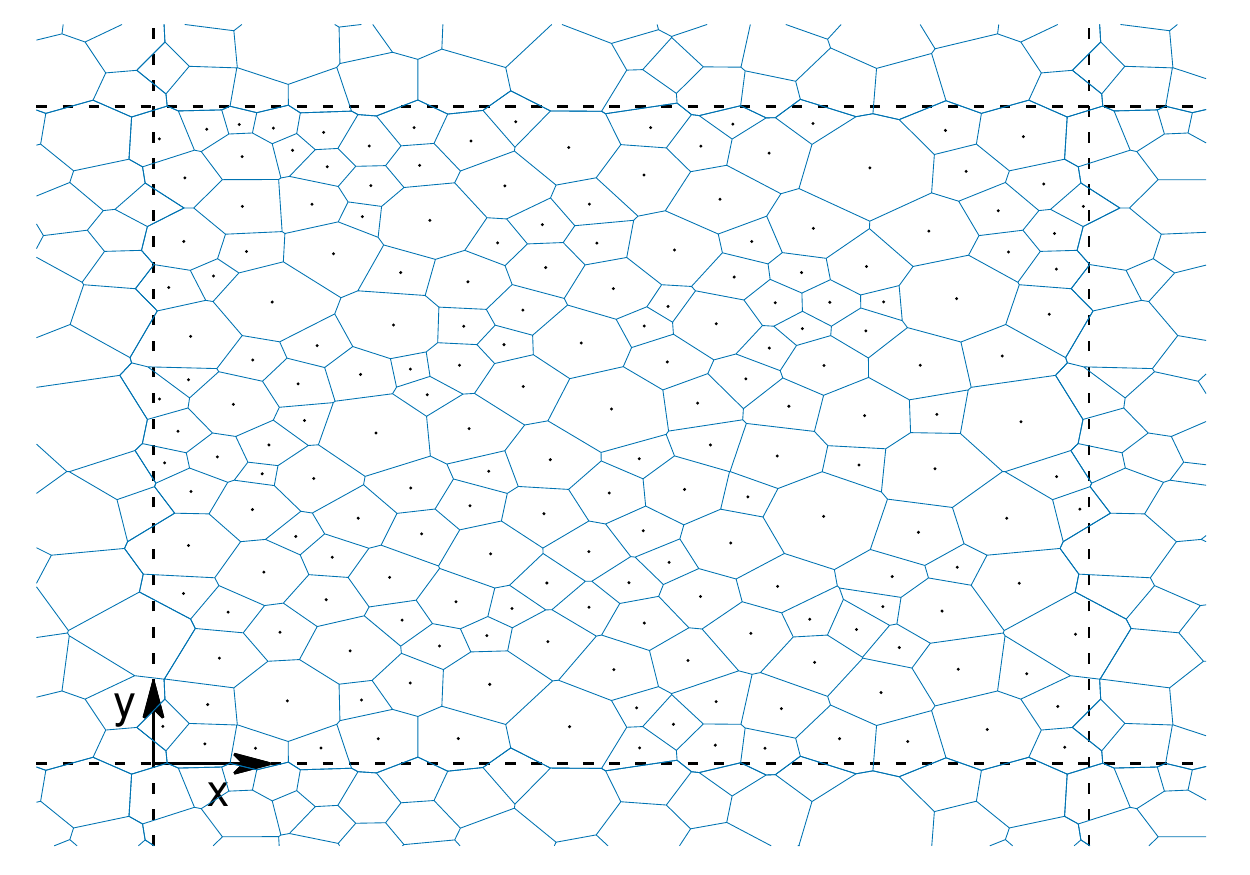}
    \caption{Example of the periodic system generated via MARS using a Laguerre-Voronoi tessellation. The grains containing points are those generated, with the dashed lines indicating the periodic repetitions.}
    \label{fig:Example_sys}
\end{figure}

For micromagnetic simulations periodic boundary conditions are used for the Voronoi tessellation in order to remove edge effects, Fig. \ref{fig:Example_sys} shows a system created via a Laguerre-Voronoi tessellation with the periodic boundaries indicated by the dashed lines. 

\subsection{System energy and effective fields}
\label{ssec:Hamiltonian}

For a magnetic film composed of $N$ individual grains, which we can consider as macrospins, the energy of the system can be written as:
\begin{equation}
    \begin{split}
        E = -&\sum_{i}V^iK^i(\mvec^{i}\cdot\hat{\mathbf{e}}^i\,)^{2} 
        -\mu_{0} \sum_i M_s^i V^i \mvec^{i}\cdot\Happvec  -  \\
        &\frac{1}{2} \sum_{nn\;ij} (M_s^i V^i M_s^j V^j) J_{ij} \mvec^{i}\cdot\mvec^{j} \\
        - &\frac{\mu_{0}}{4\pi} \sum_{i \neq j}(M_s^i V^i M_s^j V^j)\frac{3(\mvec^{i}\cdot\hat{\pmb{r}}_{ij})(\mvec^{j}\cdot\hat{\pmb{r}}_{ij})-\mvec^{i}\cdot \mvec^{j}}{r_{ij}^3}  \; ,        
    \end{split}
    \label{eq:Energy}
\end{equation}

where $V^i$ is the volume of grain $i$ with uniaxial anisotropy energy density $K^i$, easy axis direction $\hat{\mathbf{e}}^i$, bulk saturation magnetisation $M_s^i$ and normalised magnetisation vector $\mvec^{i} = \mathbf{M}^i/M_s^i$. The anisotropy energy density, saturation magnetisation and fractional exchange constant $J_{ij}$ are all temperature dependent. 
 The first term is the Zeeman energy which describes the interaction of the grains with an external applied field $\Happvec$. The second term is the anisotropy contribution to the energy which describes the preferred alignment direction for the grains magnetisation. The third term describes the exchange interaction between nearest neighbour grains which can be expressed in terms of the local exchange field $\mathbf{H}^{i}_{\mathrm{exch}}$
The fourth term is the long-range magnetostatic interaction, within the dipole approximation, which couples grains at sites $\mathbf{r_i}$ and $\mathbf{r_j}$ at a distance $\mathbf{r}_{ij} = \mathbf{r_j}-\mathbf{r_i}$ across the whole system.
The effective field acting on each grain is obtained from the energy of the system (Eq.~\ref{eq:Energy}) and is given by:

\begin{equation}
    \Heffvec^i=   \Happvec + \Hanivec^i + \mathbf{H}^{i}_{\mathrm{exch}} + \Hdmagvec^i \; .
\end{equation}
The individual terms are described in the following.

$\Hexchangevec^{i}$ describes the coupling between different grains, belonging either to the same layer or to different layers as in the case of exchange-coupled composite (ECC) media. 
In the case of a granular medium the exchange results from the intergranular medium. Although this is engineered to ensure exchange decoupling, this is not necessarily complete: in fact in the case of media for perpendicular recording the exchange, which balances the effects of the magnetostatic field, is a part of the material design. Under the reasonable assumption that the intergranular exchange is proportional to the contact area between the grains, Peng et. al. have shown that the exchange $\Hexchangevec^{i}$ is given by~\cite{Peng2011}:
\begin{equation}
    \label{eq:Exchange_field_Str}
    \Hexchangevec^{i} = \sum_{j\;\in\; neigh\;i} \Hsat J_{ij}\frac{\langle A \rangle}{A_{i}}\frac{L_{ij}}{\langle L \rangle} \mvec^{j} \; ,
\end{equation}
$J_{ij}$ is the fractional exchange constant between the adjacent grains and $L_{ij}$ is the contact length between grains \emph{i} and \emph{j}, $A_i$ is the area of grain \emph{i}, $\langle\;\rangle$ denotes the average value and \Hsat is the exchange field strength at saturation, which is generally derived from experiment. Sokalski et.al.~\cite{Sokalski2009} investigated experimentally the exchange coupling between thin layers of $CoCrPt$ separated by an oxide, finding an exchange strength which decayed experimentally with oxide layer thickness. Ellis et.al.~\cite{Ellis2017} found a similar relation using an atomistic model based on the presence of ferromagnetic impurities in the oxide layer. This study also showed the presence of higher order (biquadratic) exchange and importantly demonstrated that the intergranular exchange decayed to zero rapidly with increasing temperature, suggesting that intergranular exchange does not play a major role in the HAMR recording process. It is important to note, given the likely origin of intergranular exchange in the presence of impurity magnetic spins in the intergranular layer, that $J_{ij}$ could vary significantly. According to Peng et.al.~\cite{Peng2011} this can lead to exchange weak links which act as pinning sites and reduce the sizes of clusters arising from magnetostatic interactions.

$\Hdmagvec^{i}$ is calculated using the dipole approximation:
\begin{equation}
    \label{eq:Hdip}
    \Hdmagvec^{i} = \sum_{j\;\in\; neigh\;i} W_{ij} \mvec_{j} \; ,
\end{equation}
where $W_{ij}$ is the demagnetisation tensor of the system:

\begin{equation}
    \label{eq:W-matrix}
    W_{ij} = \frac{M_{s}V_{j}}{4\pi r_{ij}^{3}}
        \begin{bmatrix}
            \frac{3r_{ij_{x}}^2}{r_{ij}^{2}}-1 & \frac{3r_{ij_{x}}r_{ij_{y}}}{r_{ij}^{2}} & \frac{3r_{ij_{x}}r_{ij_{z}}}{r_{ij}^{2}} \\
            \frac{3r_{ij_{y}}r_{ij_{x}}}{r_{ij}^{2}} & \frac{3r_{ij_{y}}^2}{r_{ij}^{2}}-1 & \frac{3r_{ij_{y}}r_{ij_{z}}}{r_{ij}^{2}} \\
            \frac{3r_{ij_{z}}r_{ij_{x}}}{r_{ij}^{2}} & \frac{3r_{ij_{z}}r_{ij_{y}}}{r_{ij}^{2}} & \frac{3r_{ij_{z}}^2}{r_{ij}^{2}}-1
        \end{bmatrix} \; ,
\end{equation}
$V$ is the volume of the grain, $r_{ij_{\alpha}}$ is the displacement between grains \emph{i} and \emph{j}, with the subscript $\alpha=x,y,z$ denoting the component of the displacement. As $W_{ij}$ is dependent only on the position and size of the grains this matrix can be determined prior to simulation internally or via a separate external code. Improved methods to determine the W matrix are available but these produce additional computational cost. One such method is surface charge integration as discussed in \cite{Sergiu_th}. MARS is capable of accepting the W matrix as an input enabling fast implementation of alternative methods for magnetostatic determination. 

The temperature dependence of $\Hanivec^i$ is described using the following expression:

\begin{equation}
     \label{eq:Hani_temp}
    \Hanivec^i(T) = \frac{2K^i}{M_{s}^i}(m^i(T))^{\eta-1} \left( \mvec^i \cdot \hat{e}^i \right) \hat{e}^i \; ,
\end{equation}
where $\hat{e}^i$ is the unit vector aligned along the easy axis, $K^i$ is the anisotropy and $M_{s}^i$ is the zero temperature saturation magnetisation of grain \emph{i}.
Here we exploit the fact that we can express the temperature dependence of $K$ via the dependence on the magnetisation $m$ described via Callen-Callen scaling \cite{Callen1966}, which allows $K(T)$ to be expressed as:
\begin{equation}
    \label{eq:callen-callen}
    K^i(T) = K_{0}^im^i(T)^{\eta} \; .
\end{equation}
$K_0$ is the anisotropy energy density at \SI{0}{K} and $\eta$ is determined via experiment or atomistic parameterisation. Typically the exponent $\eta=3$ for uniaxial anisotropy and $\eta=2$ for 2-site anisotropy appropriate for $FePt$~\cite{Mryasov2005}.

\subsection{Dynamical and kinetic Monte-Carlo Solvers}
\label{ssec:solvers}
MARS utilises three separate solvers to describe the magnetisation, the Landau-Lifshitz-Gilbert, Landau-Lifshitz-Bloch and kinetic Monte Carlo. The LLG and LLB solvers require very short timesteps, in the region of picoseconds and femtoseconds respectively, to function and provide dynamic information about the magnetisation. The kMC is a probabilistic solver which sacrifices the dynamic information in order to enable much larger timesteps. This enables the simulation of long timescale phenomena, for example the long-term decay of written information arising from thermal activation. 
Each of the solvers presented in this work have been rigorously tested to ensure correct implementation and high accuracy. The details of these tests are provided in Supplementary Notes 1-3, with Supplementary Figures 1-7 showing comparisons between produced and expected test results.

The LLG equation of motion for each grain \emph{i}, including stochastic effects, is given by: 
\begin{equation}
    \begin{split}
    \label{eq:LLG}
    \frac{\partial{\mvec^i}}{\partial{t}} = -\frac{\gamma_e}{1+\alpha^2} &\left(\mvec^i \times (\Heffvec^i+\Hthvec^{i}) \right) \\
    &-\frac{\alpha\gamma_e}{1+\alpha^2} \mvec^i \times \left(\mvec^i \times (\Heffvec^i+\Hthvec^{i}) \right) \;.
    \end{split}
\end{equation}
The first term describes the quantum mechanical precessional motion of the magnetisation around the effective field $\Heffvec^{i}$, while the second represents the phenomenological relaxation of the magnetisation towards $\Heffvec^{i}$ \cite{Saslow2009}.
The Gilbert damping $\alpha$ couples the spin system with the environment, considered to act as the thermal bath, and determines how fast the system relaxes towards equilibrium. $\Hthvec^{i}$ is the thermal field, this stochastic field accounts for the thermally driven behaviour of the macrospin and is described by a non-correlated white noise Gaussian function.

\begin{equation}
    \label{LLG_noise}
    \begin{aligned}
        \langle H_{\mathrm{th}}^{i\alpha}(t) \rangle &= 0 \\
        \langle H_{\mathrm{th}}^{i\alpha}(t)H_{\mathrm{th}}^{j\beta}(t') \rangle &= \frac{2\alpha k_{B}T}{\gamma_{e}M_{s}V} \delta_{ij}\delta_{\alpha \beta}\delta(t-t') \; ,
    \end{aligned}
\end{equation}
where: \emph{i}, \emph{j} label the magnetisation on the respective sites; $\alpha,\beta =x,y,z$ ;  $\kB=\SI{1.381e-16}{erg K^{-1}}$ is the Boltzmann constant; $T$ is the temperature; $\delta_{\mu\gamma}$ is the Kronecker delta and $\delta(t-t^\prime)$ is the delta function. In this formulation the noise is considered to be spatially and temporally uncorrelated, i.e., white noise.

This approach works at relatively low temperatures where one can consider the grain to be fully magnetically saturated and to exhibit coherent rotation with all atomic spins remaining parallel. Under these circumstances the equation of motion need only model transverse dynamic processes and the LLG equation is valid. However, as temperature increases and approaches the Curie point this is no longer true and the Landau-Lifshitz-Bloch (LLB) equation must be used instead. The LLB introduces a longitudinal relaxation of the macrospin which accounts for the loss of magnetisation and divergence of the longitudinal susceptibility as the temperature approaches the Curie point. The LLB equation was first derived by Garanin~\cite{Garanin1990,Garanin1997}. The LLB equation is a single (macrospin) representation of the dynamical behaviour of a single grain and differs from the LLG equation in its inclusion of longitudinal relaxation of the magnetisation. Although the LLB equation parameters to be outlined below were originally derived from mean-field theory, these can be obtained from atomistic calculations. As shown by Chubykalo et. al.~\cite{Oksana_free_energy}, the LLB equation gives excellent agreement with atomistic model calculations, essentially validating its use in calculations of HAMR, which involve heating beyond the Curie temperature. The implementation of the stochastic LLB solver used by MARS follows the work of Evans et al. \cite{sLLB-II} (sLLB-II) and for each grain \emph{i} reads:
\begin{equation}
    \begin{split}
    \label{eq:sLLB-II}
    \frac{\partial{\mvec^i}}{\partial{t}} = 
    -\gamma_e \left( {\mvec^i} \times \Heffvec^i \right)
    + 
    \frac{\gamma_e\alphapara}{{m^i}^2} \left(\mvec^i \cdot \Heffvec^i \right) \mvec^i \\
    - 
    \frac{\gamma_e\alphaperp}{{m^i}^2} \left[ \mvec^i \times \left( \mvec^i \times \left(\Heffvec^i 
    + 
    \zetaperp^i\right) \right) \right] + \zetaad^i \;,
    \end{split}
\end{equation}
where $\mvec^i$ is the reduced magnetisation $\mathbf{M}^i/M(T=0)$, $m^i$ is the length of $\mvec^i$ and $\gamma_e$ is the electron gyromagnetic ratio. The first and third terms are the precessional and damping terms for the transverse component of the magnetisation, as in Eq. \ref{eq:LLG}, while the second and fourth terms are introduced to account for the longitudinal relaxation of the magnetisation with temperature. 
The stochastic LLG and LLB solvers both utilise the Heun integration scheme. The benefits of the Heun scheme are two-fold. First it provides second order accuracy in $\Delta t$ for the deterministic part, thus rendering it more numerically stable than Euler type schemes. Second, it yields the required Stratonovich solution of stochastic differential equations.
The damping of the magnetic moment is split into longitudinal \alphapara and transverse \alphaperp components given by:
\begin{equation}
    \label{eq:damping}
    \alpha_{\parallel} = \frac{2}{3}\frac{T}{T_c}\lambda\quad\text{ and }\quad \left \{
    \begin{aligned}
        \alpha_{\bot} = \lambda\left(1-\frac{T}{3T_c}\right), && \text{if } T \leq T_c \\
        \alpha_{\bot} = \alpha_{\parallel} = \frac{2}{3}\frac{T}{T_c}\lambda, && \text{otherwise.}
    \end{aligned} \right.
\end{equation}

Where $\lambda$ is the thermal bath coupling, a temperature independent phenomenological parameter, that is the same as that used in atomistic spin dynamics. The transverse damping is related to the Gilbert damping by the expression: 

\begin{equation}
    \label{Damp_relationship}
    \alpha = \frac{\alphaperp}{m} \; ,
\end{equation}

\zetaperp and \zetaad are the diffusion coefficients that account for the thermal fluctuations. The thermal noise terms are described by Gaussian functions with zero average and a variance proportional to the strength of the fluctuations:
\begin{equation}
    \label{eq:diff_coef}
    \begin{aligned}
        <\zeta_{\mathrm{ad}}^{i\alpha}(t)\zeta_{\mathrm{ad}}^{j\beta}(t^\prime)> = \frac{2|\gamma| k_{B}T\alpha_{\parallel}}{ M_sV}\delta_{ij}\delta{\alpha\beta}\delta(t-t^\prime) \\
        <\zeta_{\bot}^{i\alpha}(t)\zeta_{\bot}^{j\beta}(t^\prime)> = \frac{2k_{B}T(\alpha_{\bot}-\alpha_{\parallel})}{|\gamma| M_sV\alpha_{\bot}^2}\delta_{ij}\delta{\alpha\beta}\delta(t-t^\prime) \;.
    \end{aligned}
\end{equation}

As temperatures approach and exceed the Curie point, Eq. \ref{eq:Hani_temp} produces a fictitious longitudinal component of the anisotropy. This leads to a reduction in the longitudinal relaxation of the magnetisation as a function of temperature. To overcome this issue the anisotropy field can also be described as a function of the transverse susceptibility $\mathrm{\chi_\bot}$ \cite{Garanin1997}:

\begin{equation}
    \label{eq:Hani_chi}
        \Hanivec^{i} = \frac{-(m_x^{i} \hat{x} + m_y^{i} \hat{y})}{\chi_{\bot}} \; ,
\end{equation}

where $\mathrm{m_x^{i}}$ and $\mathrm{m_y^{i}}$ are the components of the reduced magnetisation vector and $\hat{x}$, $\hat{y}$ are the unit vector along these directions, respectively. Unlike Eq. \ref{eq:Hani_temp} this form of the anisotropy assumes that the easy axis lies along the z-axis however it is valid for all temperature ranges and is therefore the most suitable description for LLB applications. For soft materials the determination of $\mathrm{\chi_\bot}$ is extremely challenging and thus both forms of the anisotropy are available for use with the LLB solver to enable the simulation of both hard and soft materials. 

The LLB equation includes an additional field term, $\Hintragrainvec^{i}$, within the effective field. This term accounts for the exchange between the atoms within grain \emph{i}, controls the length of the magnetisation and is given by 
\begin{equation}
    \label{eq:Intragrain}
    \Hintragrainvec^{i} = \left \{
    \begin{aligned}
            \frac{1}{2\Chipara}\left(1-\frac{m^{i^{2}}}{m^2_e}\right)\mvec^{i}, && \text{if } T \leq T_c \\
           -\frac{1}{\Chipara}\left(1+\frac{3}{5}\frac{\Tc}{T-T_c}m^{i^{2}}\right)\mvec^{i}, && \text{otherwise.}
    \end{aligned} \right.
\end{equation}
Here $m^{i}$ is length of the reduced magnetisation $\mvec^{i}$ of grain \emph{i}, and 
$m_e(T)$ is the equilibrium magnetisation. The term $\Hintragrainvec$ encapsulates the new physics introduced by the LLB equation. It incorporates the longitudinal fluctuations of the magnetisation while maintaining a mean value $m_e(T)$. It is important to note that the fluctuations diverge as $\Chipara$ diverges close to $T_c$. This is responsible for the onset of the linear reversal model close to $T_c$ 
. Clearly specification of the temperature dependence of the LLB parameters is of paramount importance: a factor complicated by the effects of finite size on the magnetic properties.

\subsubsection{Atomistic parameterisation}
The granular model requires characterisation of the temperature dependence of the magnetisation, anisotropy and susceptibilities. These quantities are obtained via fitting of atomistic data, obtained using the VAMPIRE package\cite{Evans2014}. A key benefit of atomistic parameterisation is the improved accuracy of the modelled material's behaviours as well as the ability to simulate granular systems which include a segregant between the grains as is typically the case in recording media. There are two available methods for fitting the magnetisation. The first is fitted according to  $m(T)=M(T)/\Mags=(1-T/\Tc)^{\beta}/\Mags$, where \Mags is the spontaneous magnetisation and $\mathrm{\beta}$ is the critical exponent. The second is fitted via a more complex polynomial in powers of $(T-\Tc)/\Tc$:
\begin{equation}
    \label{eq:m_eq_fit}
    m(T) =
    \left \{
    \begin{aligned}
        \sum_{i=0}^{9} A_i \left( \frac{\Tc-T}{\Tc} \right)^i   &+ A_{1/2} \left( \frac{\Tc-T}{\Tc} \right)^{\frac{1}{2}} &&, \text{\;if } T < T_c \\
        \bigg[\sum_{i=1}^{2} B_i\left( \frac{T-\Tc}{\Tc} \right)^i &+ A^{-1}_0 \bigg]^{-1}  &&, \text{\;otherwise.}
    \end{aligned} \right.
\end{equation}
\begin{figure}
    \centering
    \includegraphics[width=\linewidth]{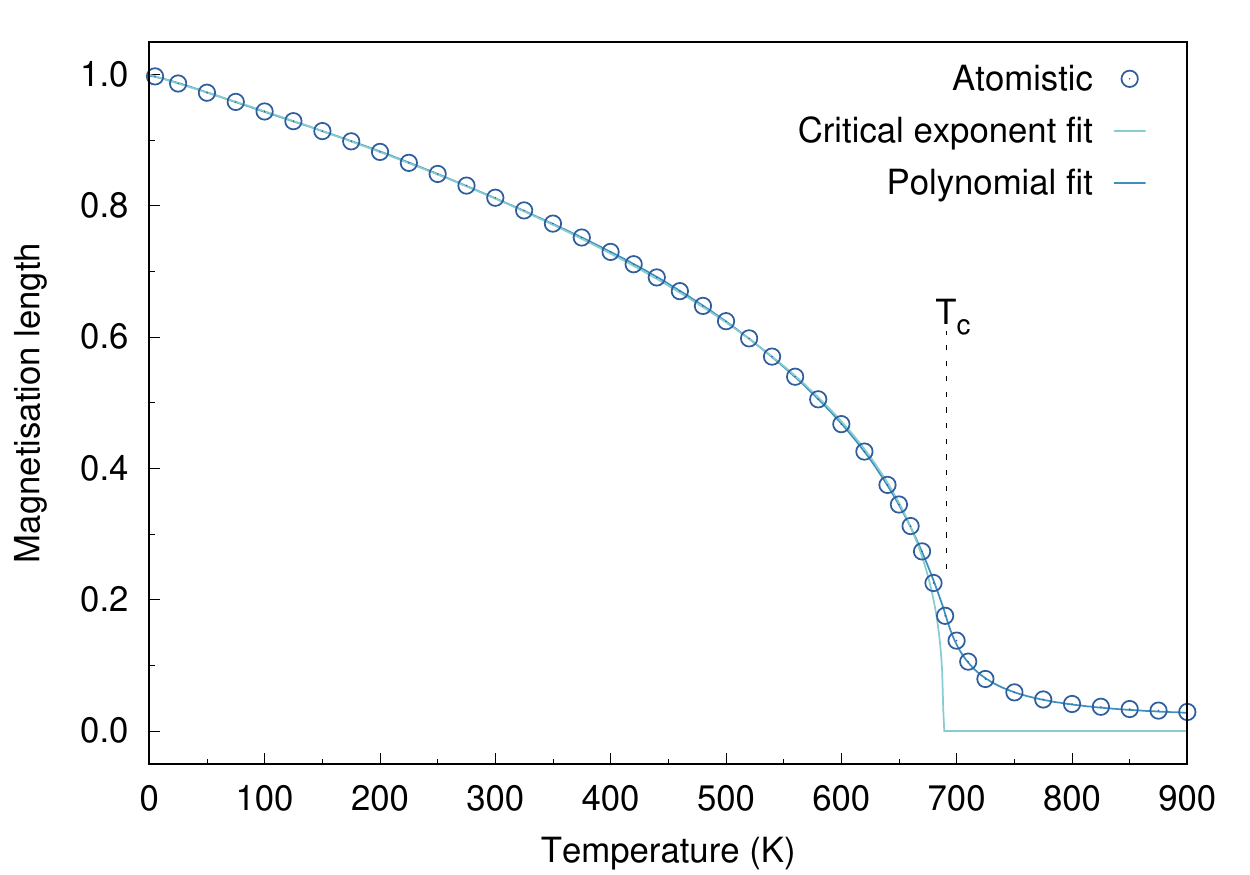}
    \caption{Comparison of the fittings of the magnetisation achieved for a \SI{5}{nm} grain via the two available methods implemented in MARS. The dots represent the atomistic data, while the lines show the fits.}
    \label{fig:Atom_meq}
\end{figure}

Both methods are capable of producing the characteristic behaviour of the temperature dependent magnetisation. A comparison of these two methods is given in Fig. \ref{fig:Atom_meq}. For bulk systems a strong criticality is expected and the critical exponent fit reproduces the sharp transition to zero magnetisation at the Curie point. However, as grain sizes decrease finite size effects become significant which cause a reduction in the criticality of the transition. The result of finite size effects is a small but non-zero magnetisation above the Curie point. The polynomial fit is capable of reproducing this behaviour and provides greater agreement with atomistic data for small grains (i.e. \SI{5}{nm}) than the critical exponent fit. 

\begin{figure}
    \centering
    \includegraphics[width=\linewidth]{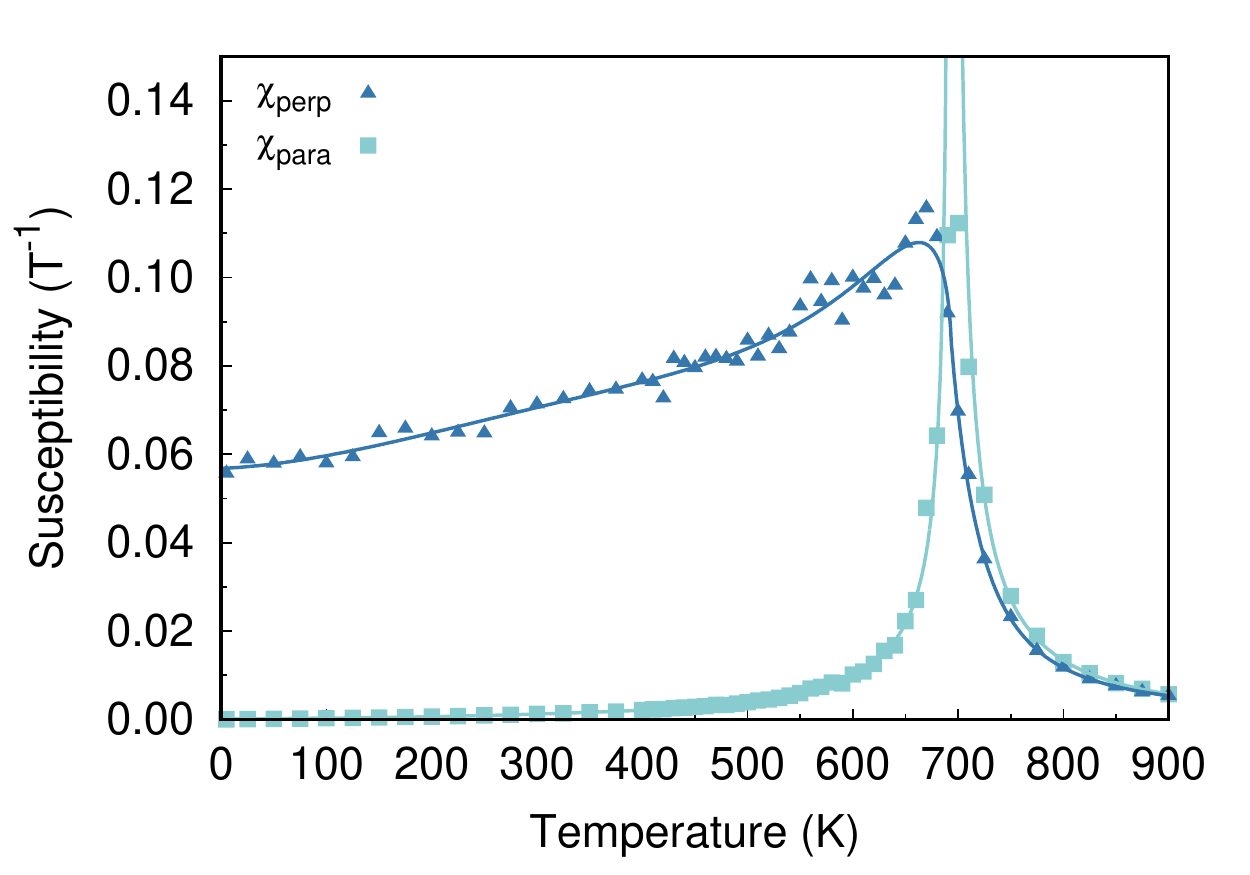}
    \caption{Fit obtained for parallel and perpendicular susceptibility using the inverse method similar to that of Ellis \cite{Matt-LLB}. The Curie point of this system is the temperature where the susceptibilities first intersect, which for this data is \SI{685.14}{K}}
    \label{fig:Atom_Chi}
\end{figure}

The susceptibility $\chi$ is a measure of the strength of the fluctuations of the magnetisation. The components of the susceptibility, according to the spin fluctuation model, can be obtained by the fluctuations of the same magnetisation components as follows \cite{Matt-LLB}:
\begin{equation}
    \Tilde{\chi}_{\mathrm{\alpha}} = 
    \frac{\smmu N}{\kB T}
    \left(
    \left\langle m_{\alpha}^2 \right\rangle - \left\langle m_{\alpha} \right\rangle^2
    \right) \, .
    \label{eq:reduced_susceptibility}
\end{equation}
Where $\Tilde{\chi}_{\mathrm{\alpha}} = {\chi}_{\mathrm{\alpha}}/\Mags V$ is the reduced susceptibility and is in units of $\mathrm{field}^{-1}$. $N$ is the number of spins in the system with magnetic moment \smmu.
Here $\left\langle m_{\alpha}\right\rangle$ is the ensemble average of the reduced magnetisation component $\alpha=x,y,z$ and longitudinal. $x,y,z$ are the spatial Cartesian components of the magnetisation, while longitudinal describes the length of the magnetisation.
\Chipara describes the strength of the fluctuations of the magnetisation component along the easy-axis direction, which for our system is $z$. \Chiperp refers to the fluctuations orthogonal to the easy axis and thus on the x-y plane.
For \Chipara and \Chiperp, we use a similar approach to Ellis \cite{Matt-LLB} and we fit the inverse of the susceptibility $1/\Chiparaperp$: 
\begin{equation}
    \label{eq:susceptibility}
    \frac{1}{\Chiparaperp} = 
    \left \{
    \begin{aligned}
        \sum_{i=0}^{9} C_i \left( \frac{\Tc-T}{\Tc} \right)^i &+ C_{1/2} \left( \frac{\Tc-T}{\Tc} \right)^{\frac{1}{2}} &&, \text{\;if } T < T_c \\
        \sum_{i=0}^{4} D_i \left( \frac{T-\Tc}{\Tc} \right)^i &  &&, \text{\;otherwise.}
    \end{aligned} \right.
\end{equation}
Where $C_i$ and $D_i$ are the fitting parameters and \Tc is the Curie point, obtained by determining the temperature at which the susceptibilities intersect. Fig. \ref{fig:Atom_Chi} shows the susceptibilities and fits obtained from atomistically parameterised FePt, the Curie point of this system is \SI{685.14}{K}.

Low anisotropy systems and systems of reduced dimensions cannot retain the alignment of the magnetisation along the easy axis up to \Tc. In such cases \Chipara is a mix of the spatial components and becomes difficult to determine.
A workaround is to avoid calculating \Chipara directly and to obtain \Chipara from the longitudinal susceptibility \Chilong, following the discussion presented in \cite{Vogler2014_LLB}.
Unfortunately a similar method cannot be used for \Chiperp making it difficult to determine the anisotropy for soft systems when the anisotropy field is given by Eq. \ref{eq:Hani_chi}.

\begin{figure}
    \centering
    \includegraphics[width=\linewidth]{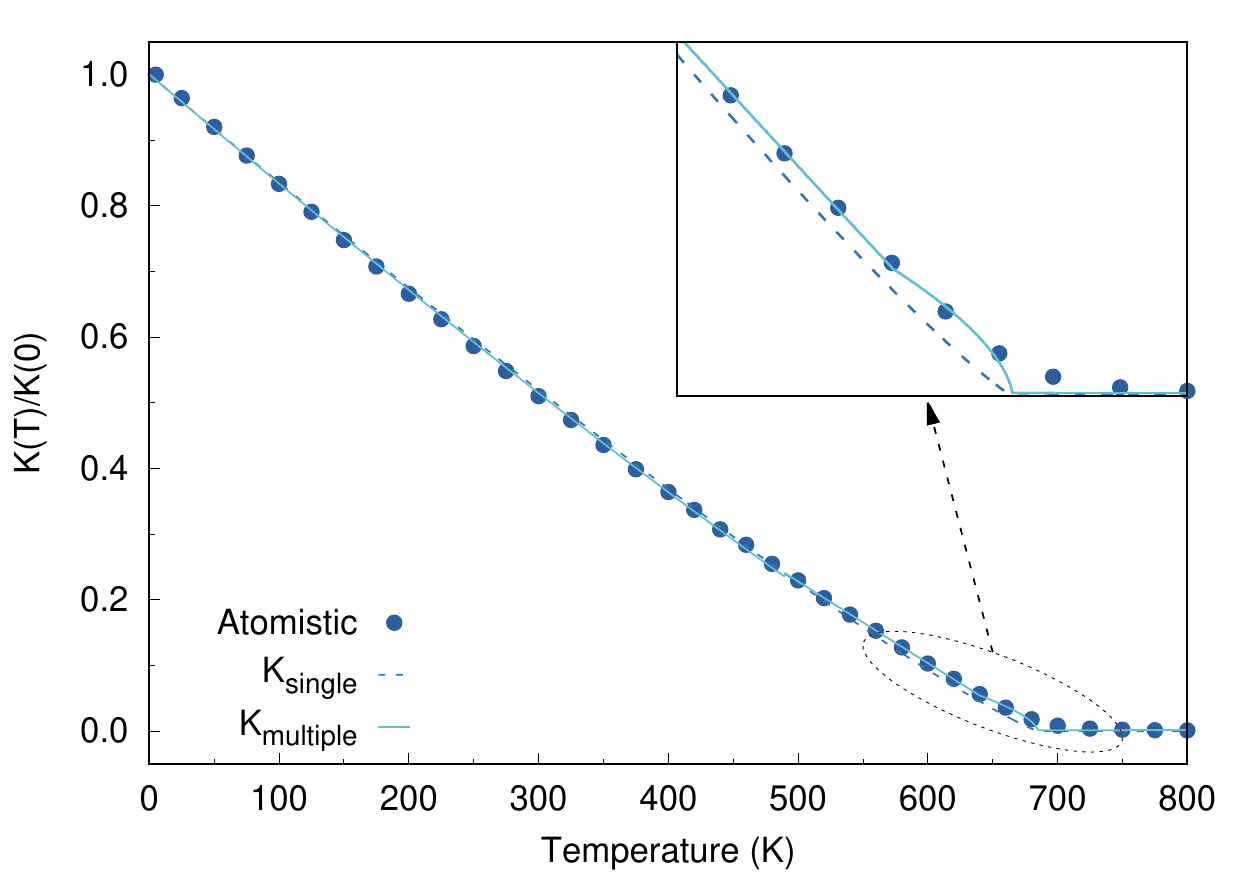}
    \caption{Comparison of the fit achieved via the standard (single) and extended (multiple) Callen-Callen methods. The points represent the atomistic simulation data while the lines show the fits used in MARS. The extended Callen-Callen is able to provide an improved fit overall and especially near the Curie point as shown in the inset.}
    \label{fig:Atom_Callen-Callen}
\end{figure}

Alternatively, if the anisotropy field is described as in Eq. \ref{eq:Hani_temp}, the reduced anisotropy is given by $k(T)=K(T)/K_0=m(T)^{\gamma}$, as discussed by Callen-Callen \cite{Callen1966}. 
MARS implements both a standard Callen-Callen fitting and an extended version. The extended version utilises three temperature regions each with their own fit parameters such that there are no discontinuities. This extended fitting method enables greater accuracy in the reproduction of the anisotropy as a function of temperature. 
This approach should provide more useful results in the case of soft materials, where extracting \Chiperp can prove difficult. Fig. \ref{fig:Atom_Callen-Callen} is a comparison of the fits obtained using the standard and extended Callen-Callen fitting methods. 
Once all these parameters are determined, the granular model is fully parameterised regarding the material properties.

\subsubsection{Kinetic Monte-Carlo solver}
Finally we turn to the solver for long-timescale simulations: the kinetic Monte Carlo (kMC) solver. 
In the kMC approach as given in~\cite{Chantrell2001a}, the switching probability is dependent on the measuring time $t_m$ as follows,
\begin{equation}
    P_t=1-e^{-t_m/\tau} \; ,
    \label{prob}
\end{equation}
where the relaxation time $\tau$ is given by the Arrhenius-N\'{e}el law~\cite{NEEL}
\begin{equation}
    \tau^{-1} = 
    f_{0}\Big(\exp\Big[-\frac{\Delta E_{12}}{k_{B}T}\Big]+\exp\Big[-\frac{\Delta E_{21}}{k_{B}T}\Big]\Big) \; ,
\end{equation}
where $f_{0}$ is the attempt frequency, usually assumed around $10^{9}$ s$^{-1}$ for these systems, and $\Delta E_{12,21}$ are the energy barriers for switching between states.  
Typically, large energy barriers of $>60k_{B}T$ are required in order to ensure long-term thermal stability of written  bits.
$\tau^{-1}$ is given by $\tau^{-1}=\tau^{-1}_{12}+\tau^{-1}_{21}$.
To model the physical effect of the easy axis dispersion, easy axes are chosen randomly within a Gaussian dispersion of angle about the normal. The total energy barrier including the effect of anisotropy dispersion can be written as

\begin{equation}
    \Delta E({\bf H}_{\mathrm T}, \psi)=K_{u}V[1-{\bf H}_{\mathrm T}/g(\psi)]^{\kappa(\psi)} \; , \label{Eb}
\end{equation}
where ${\bf H}_{\mathrm T}$ the total effective field, $g(\psi)=[\cos ^{2/3}\psi+\sin^{2/3} \psi]^{-3/2}$ and $\kappa(\psi)=0.86+1.14g(\psi)$ are the numerical approximations given by Pfeiffer~\cite{Pfeiffer1990}. 
The kMC is capable of simulations on the timescale of years and is valid for systems where the energy barrier is much larger than the thermal energy.  In order to function, the kMC requires calculation of the magnetisation states corresponding to the energy minima along with the energy barrier separating the two states. Stationary states are found by solution of the Stoner-Wohlfarth model of coherent rotation~\cite{Stoner-Wohlfarth} for which the free energy is given by
\begin{equation}
    E = K_uV(\hat{\mathbf e} \cdot \hat{\mathbf m})^2 -\mu \mathbf H \cdot \hat{\mathbf m} \; .
\end{equation}
The final step  \cite{Chantrell2001a} is to ensure that, after switching, the populations of the energy minima obey Boltzmann statistics. This approach leads to the condition that,
if the reversal transition is allowed, the moment is then assigned to either energy minimum
with a probability,
\begin{equation}
    p=e^{-E_{i}}/(e^{-E_{1}}+e^{-E_{2}}) \; ,
    \label{eq:prob}
\end{equation}
with $i=1,2$  labelling the minima, thereby ensuring that the population of the two states obeys the Boltzmann distribution in thermal equilibrium \cite{Chantrell2001a}. This is important in order to include the `backswitching' mechanism which leads to dc or `remanence' noise.
Taking account of both distributions the transition probability is determined using:

\begin{equation}
    \label{eq:kMC_prob}
    P_2 = \bigg(1-\exp\Big[-\frac{t_m}{\tau}\Big]\bigg)\bigg(1+\exp\Big[-\frac{(E_2-E_1)}{k_{B}T}\Big]\bigg)^{-1} \; .
\end{equation}

Where $P_2$ is the probability of the magnetic moment jumping to the second minima and $\Delta E$ is the energy barrier separating the two minima. To determine if a switching event occurs a random number between zero and unity is generated and compared to $P_2$. If it is less than $P_2$ the magnetic moment orientation is assigned corresponding to the second minimum otherwise it is assigned to the first minimum.
$t_m$ is the measurement time. During the measurement time the external properties such as magnetic field and temperature are assumed constant, such that Eq. \ref{eq:kMC_prob} can be applied.  

\subsection{Curie temperature dispersion}
The HAMR process involves heating through $T_c$ which, as a result, becomes an important material parameter. More particularly, simulations by Li and Jhu~\cite{Li2013,Zhu2014} have shown that the dispersion of $T_c$ is a serious limitation for the ultimate storage density achievable for HAMR. Here we consider an irreducible contribution to the dispersion of $T_c$ which arises directly from the diameter dispersion. It is well known that finite size effects lead to a reduction of $T_c$, demonstrated experimentally for $FePt$ by Rong et.al.~\cite{Rong2006}. A theoretical investigation based on an atomistic model by Hovorka et. al.~\cite{Hovorka2012a} showed that the variation $M(T)$ was well described by the finite size scaling law
\begin{equation}
    \label{Eq:FSSL}
    T_{c}(D) = T_{c}^{\infty}(1-(d_{0}/D)^{\lambda}) \; ,
\end{equation}
where $D$ is the grain diameter, $\lambda$ is the so-called phenomenological shift exponent and
$d_0$ is the microscopic length scale close to the dimension of a unit cell of the lattice structure. The exponent $\lambda$ is related to the correlation length
universal critical exponent $\nu$ and it is
expected that $\lambda=\nu^{-1}$. However, small grains can exhibit departure from universality so we prefer the form of Eq.~\ref{Eq:FSSL} as a functional form to represent the diameter dependence of $T_c$. Clearly a dispersion of diameter maps onto the dispersion of $T_c$. Assuming a lognormal distribution of $D$, with logarithmic mean $D_m$ and variance $\sigma_D^2$ it has been shown~\cite{Hovorka2012a} that
the dispersion of $T_c$ is given by the distribution function
\begin{equation}
    \label{tcdistr}
    f_T(\Delta T_c) = \frac{1}{\sqrt{2\pi}\Delta T_c\sigma_T}
                                 \exp{\left(-\frac{(\ln\Delta T_c- T)^2}{2\sigma_T^2}\right)} \; ,
\end{equation}
with $\Delta T_c = T_c^\infty-T_c$. Eq.\eqref{tcdistr} is a lognormal distribution function with logarithmic mean  $T_m = \lambda(\ln(d_0(T_c^{\infty})^{1/\lambda})- D_m)$ and variance $\sigma_T^2 = \lambda^2\sigma_D^2$. Through Eq.~\ref{Eq:FSSL}, with $d_0$, $\lambda$ and $T_c^{\infty}$ determined either from experiment or atomistic model calculations, a $T_c$ value can be assigned to an individual grain and 
Eq.~\ref{tcdistr} used to calculate the standard deviation of the $T_c$ dispersion.

\section{Simulations}
\label{sec:Simulations}

This section presents a demonstration of some of the bundled simulation types available in MARS at release. These simulations serve to show the capability of MARS and provide some of the most common simulation types. The first example simulation is the writing of a pseudorandom bit sequence (PRBS) pattern for HAMR. The second example is the determination of switching probabilities for characterising Curie point dispersion via comparison with experimental data. Finally an example of FMR simulations for use in determining system damping is provided.

\subsection{Writing and reading processes for heat assisted magnetic recording}

Development of improved heat assisted magnetic recording requires investigations into the thermal reversal of the grains along with the effect of bit spacing on data stability and writing performance. MARS contains three separate HAMR focused simulations. The first applies a temporally dependent laser and field profile onto a single bit, allowing  investigations of  thermal reversal properties of materials. The second simulation, consists of a bit within a surrounding system of grains, utilising a temporal and spatially dependent laser and field profile, allowing for investigation into the influence of the written bit on any surrounding grains. The third simulation models realistic data writing, via a square-wave or user specified binary sequence. This writing can occur for single or multiple tracks. Using this third simulation, comprehensive investigations into the entire HAMR writing process can be achieved.

Read back of a system can also be simulated in MARS. The system is first discretised into \SI{1}{nm} pixels, this enables more precise measurement of the magnetisation and produces a more realistic read back signal. A read head is then placed at the edge of the track. This head is scanned along the track and the magnetisation detected within the head is averaged and recorded as a function of down track position. In order to investigate data decay over time the simulation utilises the kMC solver to simulate the system for years and performs read back at specified intervals. 

\begin{figure*}[ht]
    \centering
    \includegraphics[width=1.0\textwidth]{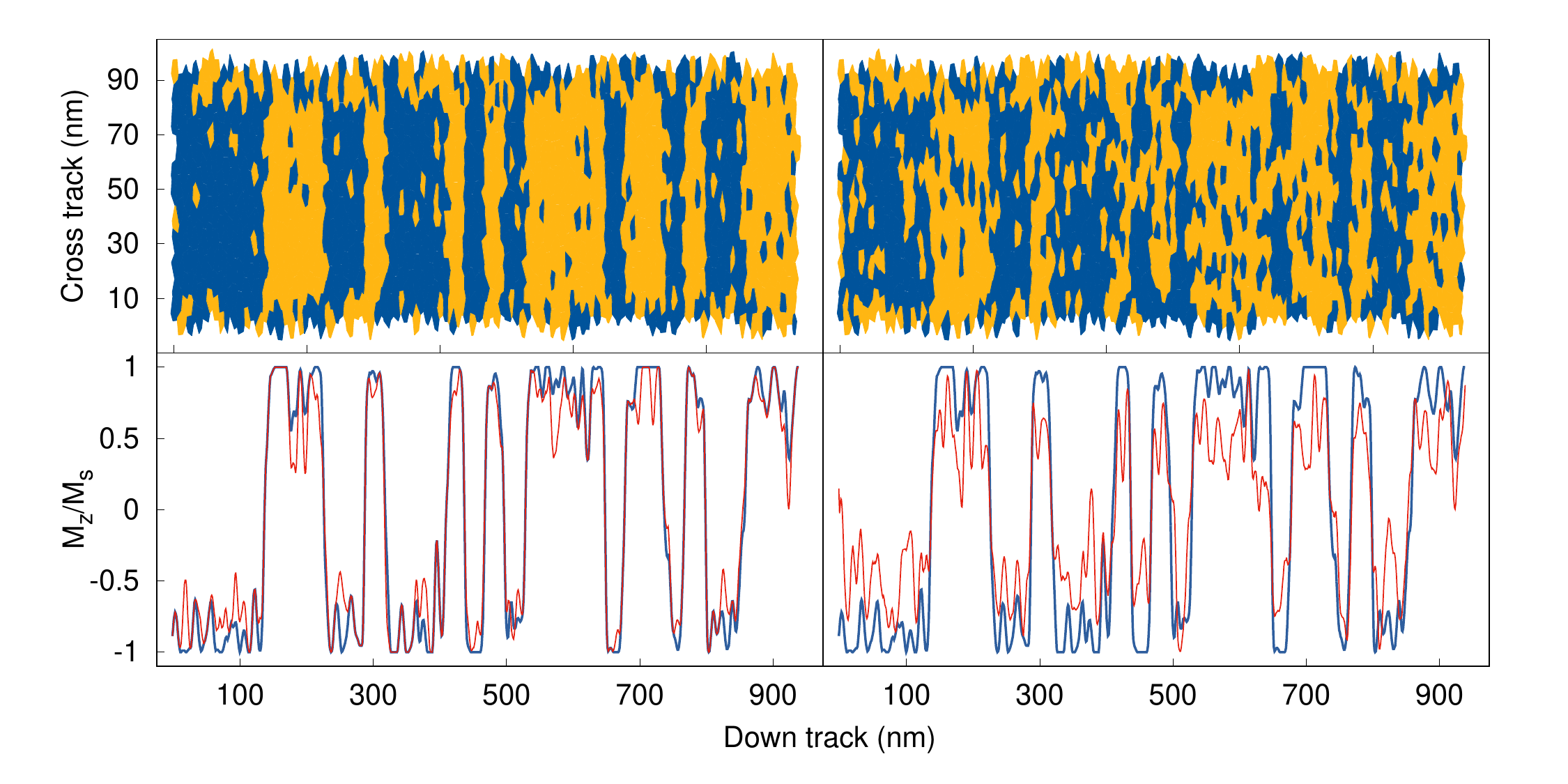}
    \caption{HAMR writing and reading simulation for $KV/kT=79$ on the left and $KV/kT=62$ on the right. The top graph shows the magnetisation of the grains after the PRBS was written via the LLB solver. The bottom graph shows the signal obtained via read back immediately after writing (blue) and after a ten year time evolution (red) via the kMC solver. 
    }
    \label{fig:HAMR_WR_ALL}
\end{figure*}

\begin{figure}[p]
    \centering
    \includegraphics[width=\linewidth]{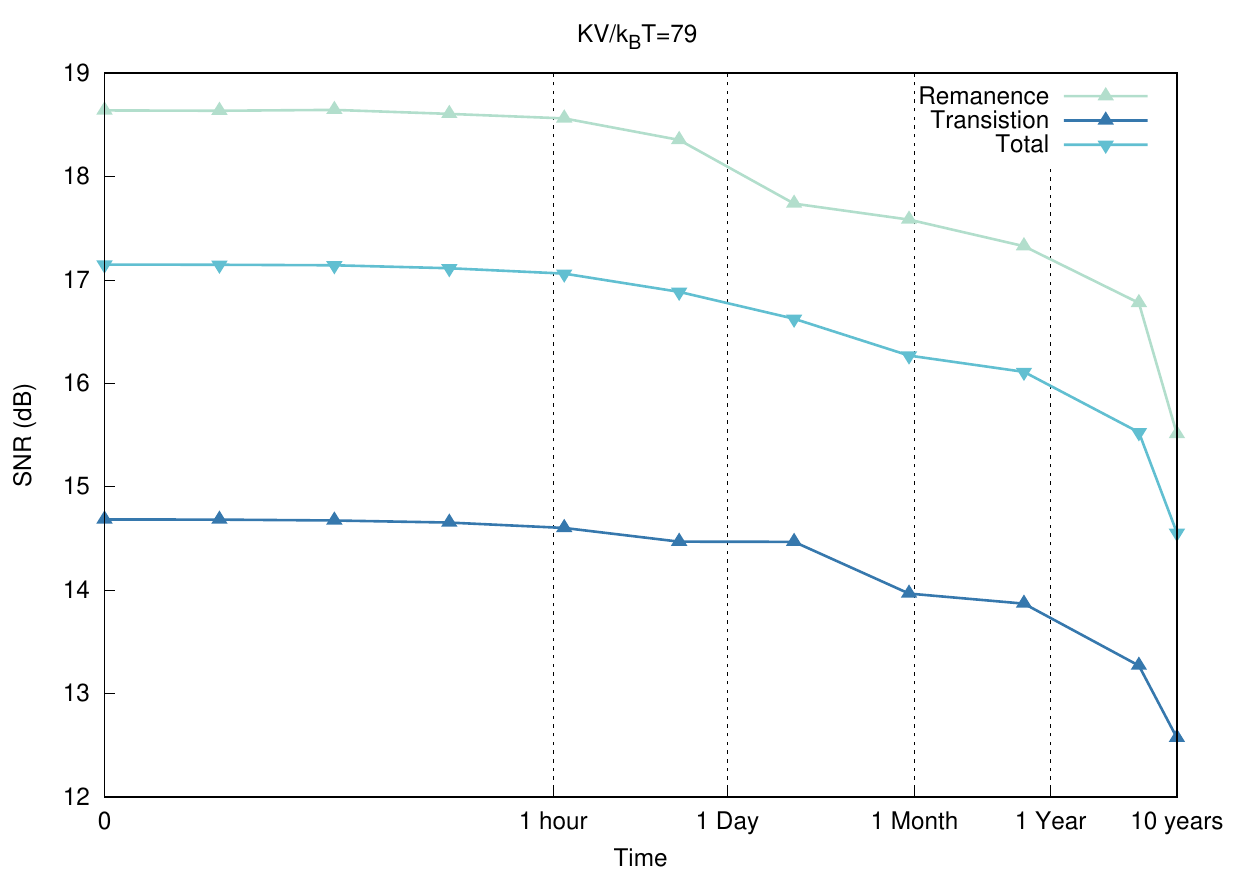}
    \caption{SNR obtained from multiple read back simulations performed for a HAMR system with $KV/k_BT=79$.}
    \label{fig:SNR}
\end{figure}

Fig. \ref{fig:HAMR_WR_ALL} shows the output obtained via the realistic HAMR writing and reading simulation available in MARS. For this simulation two systems with energy barriers of $KV/k_BT=79\;\mathrm{and}\;62$ were used. Each system consisted of 1,300 grains with a lognormal grain size distribution with an average diameter of \SI{8}{nm}. The material parameters used were $\Mags=$ \SI{1051.65}{emu/cm^3}, $T_c=$ \SI{693.5}{K}, $\lambda=$ \SI{0.1}{} and $K_u=$ \SI{9.2e7}{erg/cm^3} and $K_u=$ \SI{7.1e7}{erg/cm^3} respectively. A 31-bit PRBS was written to the systems. The sequence used was 1111100011011101010000100101100. Both systems were then evolved over time for ten years and read back to generate a second read back signal. 
This simulation was performed 100 times with different random seeds, the read back signal was then obtained for each simulation in order to obtain the spatial, i.e. of the medium, signal-to-noise ratio (SNR) as a function of time. There are two contributions to the spatial SNR: transition and remanence \cite{Victoria2013}. These SNRs describe the noise present at and away from the bit transitions respectively. The latter is caused by grains with $KV/k_BT$ such that switch back, whereas the former is a measure of how good a transition between bits is written. Transition noise is the noise expected to be dominant in high density media due to the reduce dimensions the bit size.
The calculation of SNR is currently not included in the MARS software package, even though there is the plan to include it in future releases. 
The extraction of SNR is based on the ensemble wave form analysis developed by Seagate \cite{Hernandez2016,Hernandez2017}, a method that allows to separate and extract the different noise components: transition and remanence. The approach proceeds as follows: a track is written multiple times, 10 times in the present case, with each track read back once only since reader noise is not included. The signals are first synchronised via cross-correlation and afterwards the average ``noise-free'' signal is obtained by averaging over the signal of the 10 written tracks. Then, the total spatial noise is calculated as the variance of the average ``noise-free'' signal. To extract the remanence and transition noise appropriate windowing functions are applied to the total spatial noise.  Eventually, the SNRs are calculated as the $10\log_{10}$ of the ratio between the total signal power and the respective noise power. 
Fig. \ref{fig:SNR} shows the SNR components for the high energy barrier system ($KV/k_BT=79$), and as expected the transition noise is the bottleneck.
 
\subsection{HAMR switching probability}
The heating in HAMR systems occurs in a narrow temperature region near $T_c$, thus the performance of HAMR is highly dependent on the $T_c$ dispersion of the recording medium \cite{Hai2014,Li2013,Zhu2014}.
To determine the $T_c$ dispersion experimentally the procedure requires measurement of the thermo-remanence magnetisation as a function of the applied laser pulse peak power \cite{Tc2014}. The application of this laser pulse occurs over the long timescale thus LLB simulations are extremely computationally expensive, to the point of impracticability. MARS overcomes this problem by utilising the multi-timescale nature of the code. The solver used by the code is determined automatically based on the laser application time: if this time is of the order of microseconds or greater then the kMC is used otherwise the LLB is used. The heating and cooling phases require dynamic information to be modelled and hence are always performed via the LLB solver. The ability for MARS to automatically select the most suitable solver allows for the simulations to be run in batch jobs without specification of the solver or editing of the source code. 

\begin{figure}[p]
    \centering
    \includegraphics[width=\linewidth]{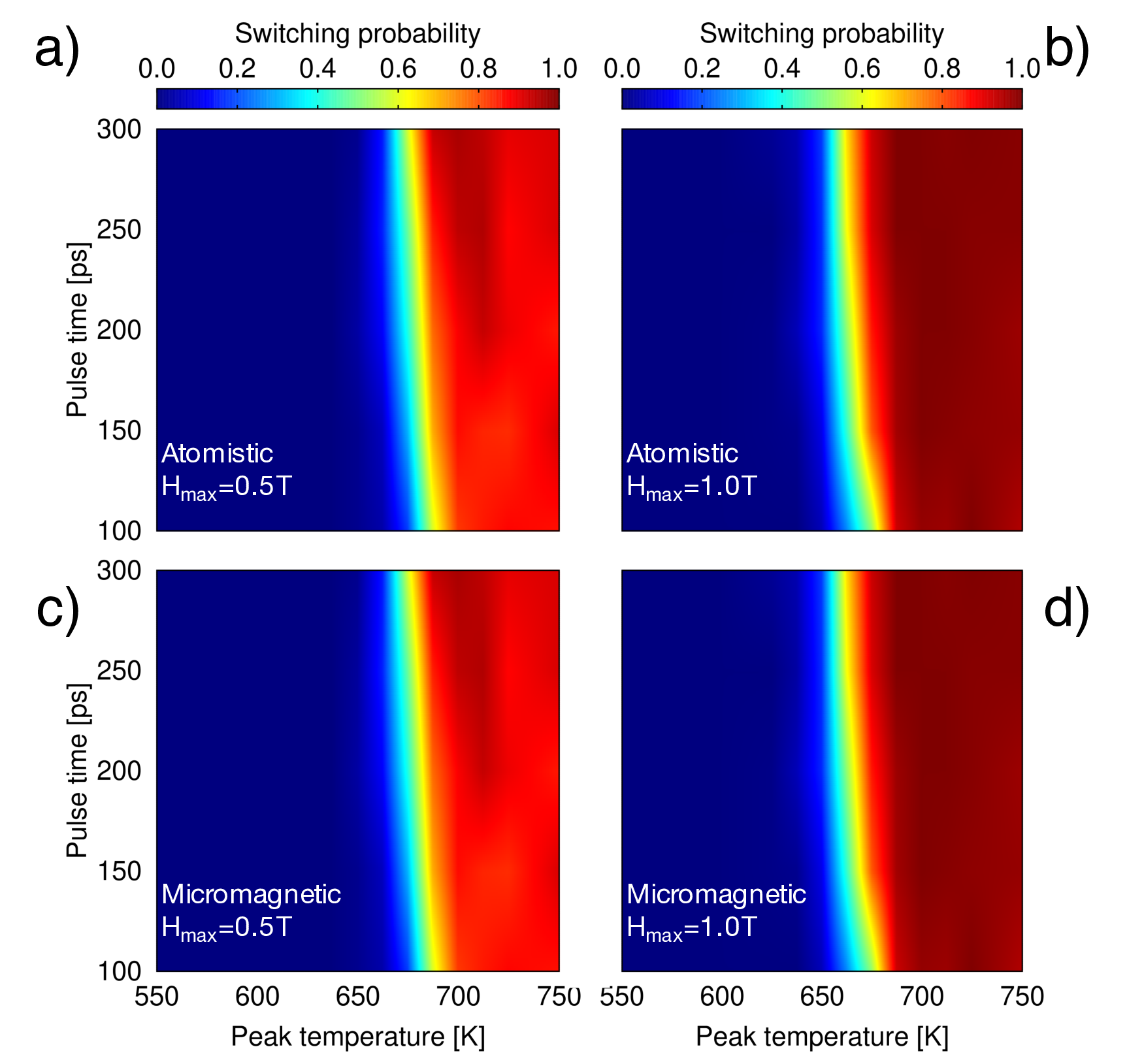}
    \caption{Switching probability as a function of peak temperature for various pulse lengths for both micromagnetic and atomistic simulations. Adapted from Ref. \cite{meo2020magnetisation}.}
    \label{fig:Andrea_SW}
\end{figure}
\begin{figure}[]
    \centering
    \includegraphics[width=\linewidth]{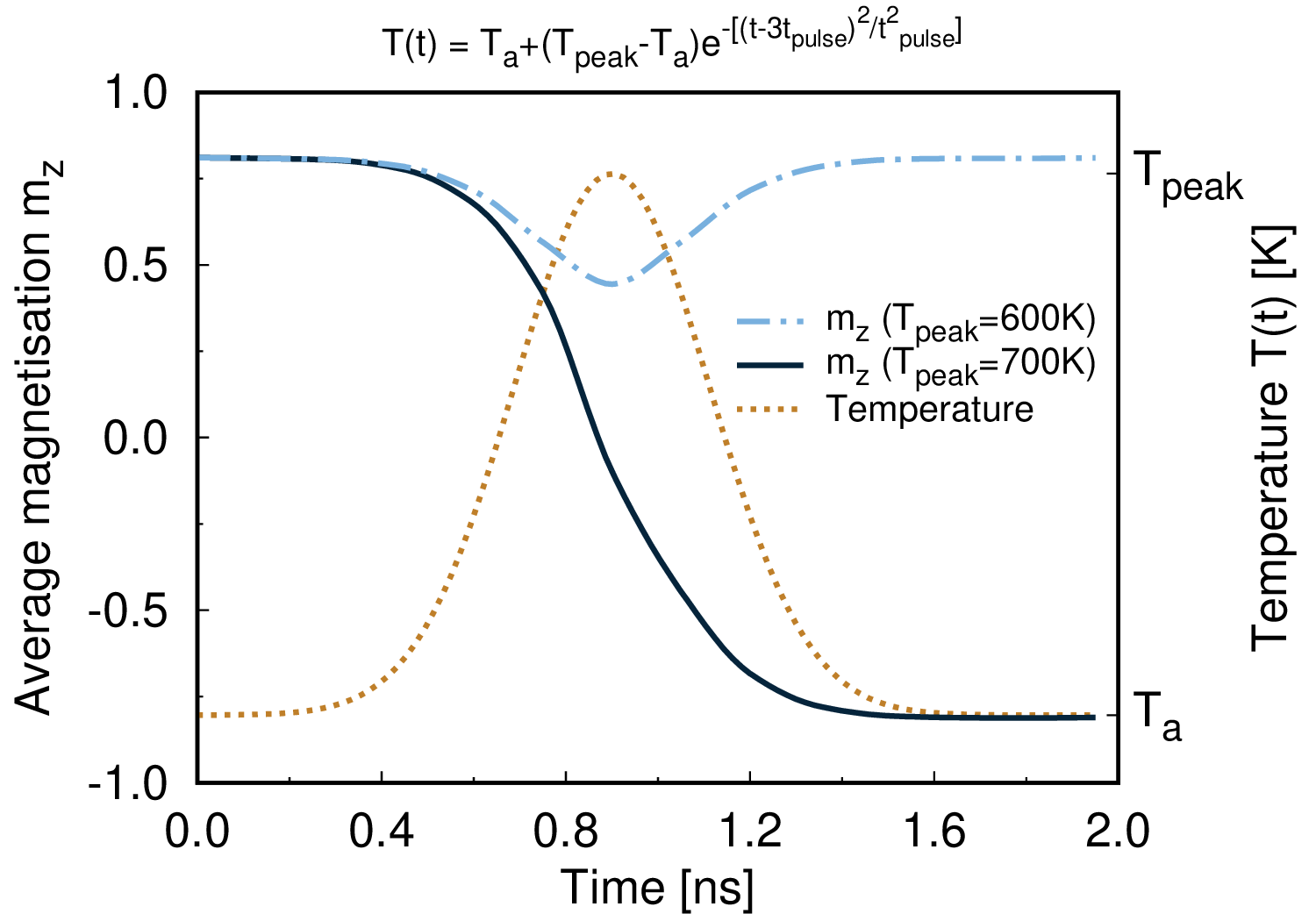}
    \caption{Time dependence of the average magnetisation for 100 FePt grains, obtained with the LLB solver, for a heat pulse with $\tpulse=\SI{300}{ps}$ and \Tpeak \SI{600}{K} (dot-and-dash blue) and \SI{700}{K} (solid black) under an applied field of \SI{-1}{T} along the $z$--direction. The brown dotted line shows the time dependence of the temperature pulse.}
    \label{fig:Mz_and_Temp_vs_time}
\end{figure}

Fig. \ref{fig:Andrea_SW}(c,d) are the results of a thermoremanence simulations used to investigate the switching probability of single FePt grains as a function of peak temperature for HAMR-like heat pulses for various pulse lengths and under different applied fields. The considered pulse length falls within the few nanoseconds regime, thus the LLB solver has been employed in these simulations. These results are compared with the switching probabilities obtained by performing atomistic simulations for the same system and setup, presented in panels (a,b). The agreement between the atomistic parameterised LLB and atomistic simulations, which have also been used for the parameterisation, proves the ability of MARS in describing such processes, as discussed in more detail in Ref. \cite{meo2020magnetisation}.
Fig. \ref{fig:Mz_and_Temp_vs_time} presents an example of the average magnetisation dynamics for 100 FePt grains under the application of an external \SI{-1}{T} field along the $z$--axis, and a \SI{1.8}{ns} heat pulse with peak temperature $\Tpeak=600\, , \SI{700}{K}$. The dashed brown line shows the Gaussian profile used to model the time dependence of the heat pulse:
\begin{equation}
\label{eq:heat_pulse_profile}
T(t)=\Tamb+(\Tpeak-\Tamb)\exp{[-(t-3\tpulse)^2 / \tpulse^2}] \; ,
\end{equation} 
where \tpulse corresponds to $1/6$ of the total pulse length and \Tamb is the ambient temperature, the temperature at which the grains are in absence of the heat pulse, \SI{300}{K} in this case. For low \Tpeak and relatively fast pulses the 
magnetisation reversal of the grains cannot be achieved, however, despite this a partial reversal can be observed when the temperature approaches \Tpeak. However, as the grains cool down the magnetisation is restored along the initial direction. On the other hand, when \Tpeak approaches \Tc of the grains, all grains are reversed.

\subsection{Ferromagnetic resonance}

\begin{figure}[ht]
    \centering
    \includegraphics[width=\linewidth]{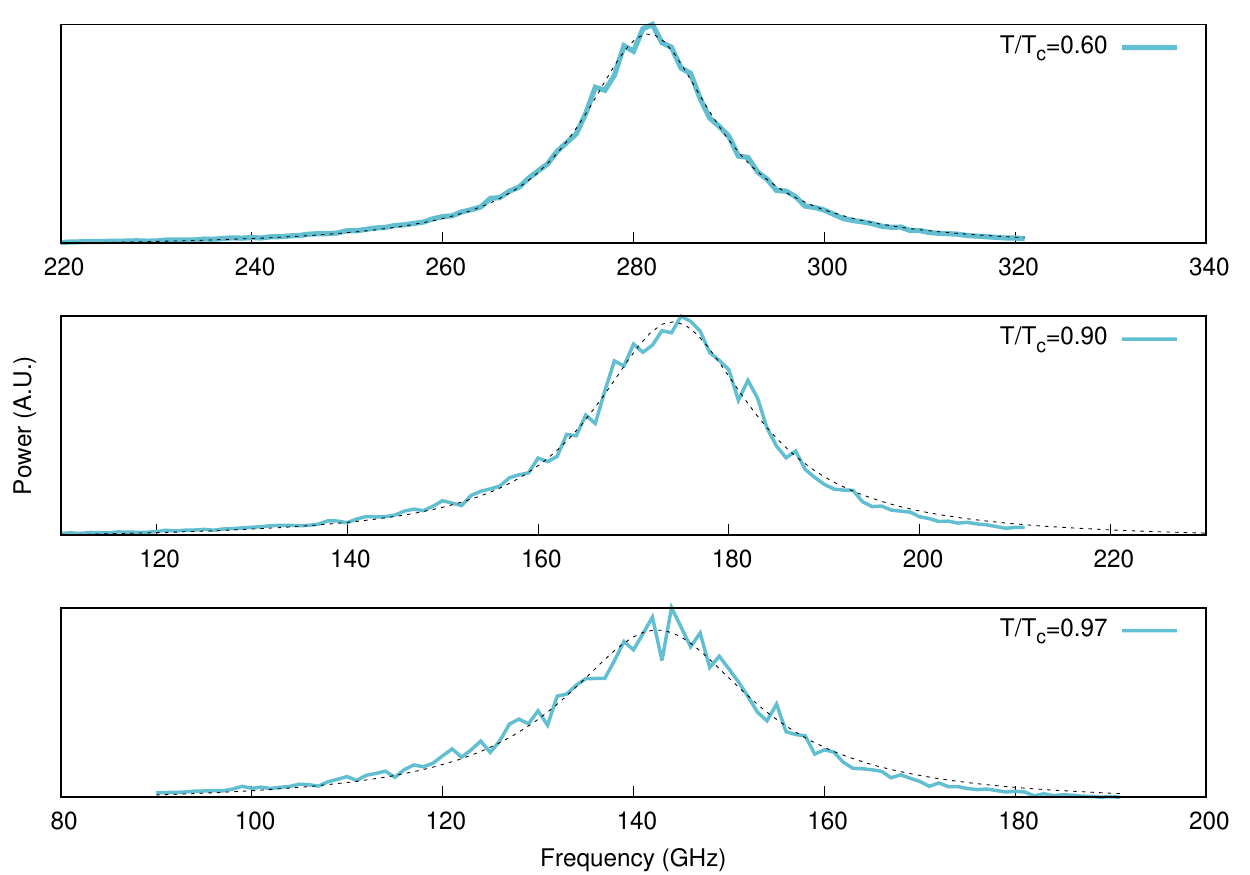}
    \caption{Frequency swept FMR for FePt at various temperature values. The Power spectra have been fitted via a Lorentzian (dashed) to obtain the system resonant frequency and damping.}
    \label{fig:FMR}
\end{figure}

\begin{figure}[ht]
    \centering
    \includegraphics[width=\linewidth]{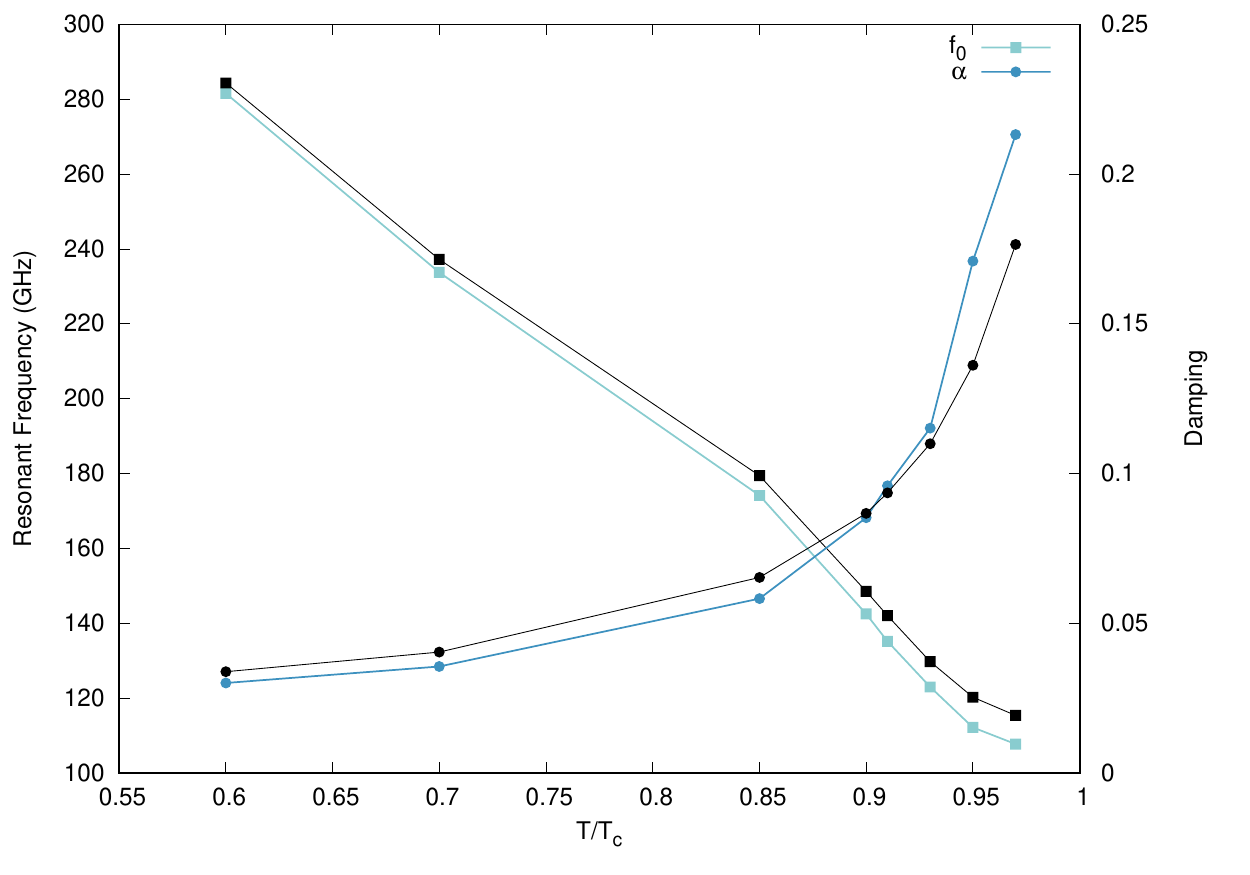}
    \caption{Damping and resonant frequency as a function of temperature for a micromagnetic FePt system parameterised using $\mathrm{5nm\;\times\;5nm\;\times\;10nm}$ FePt atomistic simulations compared to the expected analytical values (black lines).}
    \label{fig:FMRT}
\end{figure}

Ferromagnetic resonance (FMR) is an important technique used for the measurement of magnetic properties for bulk and thin-film magnetic media \cite{hashmi2015reference}. FMR simulations enable investigation into the dynamic properties of a material such as the damping \cite{yalcin2013ferromagnetic} as well as the static properties such as saturation magnetisation and uniaxial anisotropy \cite{eriksson2017atomistic}. MARS provides a simple system to perform frequency and field swept FMR simulations over a range of temperatures via its LLG and LLB solvers. Fig. \ref{fig:FMR} shows the power spectrum obtained via FFT vs. the applied field frequency obtained via FMR simulations of a single system at various temperatures. The data have been fitted to a Lorentzian function:
\begin{equation}
    L(x) = \frac{A}{\pi}\frac{0.5w}{(x-f_0)^2+(0.5w)^2},\;\;\; \alpha=\frac{0.5w}{f_0} \; ,
    \label{eq:Lorentz}
\end{equation}
where $\mathrm{A}$ is the amplitude, $\mathrm{w}$ is the full width at half maximum, $\mathrm{\alpha}$ is the damping parameter and $\mathrm{f_0}$ is the resonant frequency. The resonance frequency can be determined via the Kittel formula \cite{kittel}.

\begin{equation}
    f_{0} =  \frac{\gamma}{2\pi}(B+\mu_0 H_{k}) \; ,
    \label{eq:Kittel}
\end{equation}

where $\mathrm{B}$ is the in-plane FMR field amplitude. From this fitting the system's damping parameter and resonant frequency are extracted. Fig. \ref{fig:FMRT} shows the damping and resonant frequency extracted as a function of system temperature, there is very good agreement between the extracted results and the analytical values.

\section{Conclusion}

We have developed an open-source multi-timescale micromagnetic code (MARS) for simulating granular thin films. The primary focus of MARS is to enable detailed modelling and simulation of state of the art and future magnetic recording systems covering both the short timescale writing/reading processes and long timescale data storage. The functionality of MARS is provided via the development and inclusion of three micromagnetic solvers: the Landau-Lifshitz-Gilbert (LLG), Landau-Lifshitz-Bloch (LLB) and kinetic Monte Carlo (kMC). Short timescale simulations are possible via the LLG and LLB dynamic solvers. The LLB enables the simulation of systems up to and exceeding their Curie points, which is crucial for recording system such as HAMR. The long timescale simulations are performed via the kMC stochastic solver. The combination of these solvers allows for complex multi-timescale simulations to be developed and performed.

MARS has also been developed for use in material characterisation to aid research into the development and optimisation of recording media materials. The implementation of atomistic parameterisation enables highly accurate material descriptions within MARS and produces very good agreement between results obtained using MARS and those obtained atomistically. Thus MARS is highly useful for bridging the gap between atomistic simulation and real world experimentation. Furthermore to ensure an accurate description of granular media the numerous methods of Voronoi construction have been investigated and compared resulting in the Laguerre-Voronoi method being implemented in order to ensure realistic grain shapes as well as lognormal grain size distributions. Detailed descriptions of the models incorporated in MARS have been provided along with the various methods used to model temperature dependent material parameters. 

Finally we have provided example results obtained via MARS for numerous published or on-going studies to show the versatility and capabilities of MARS. These studies cover the entire range of HAMR development starting from materials characterisation and optimisation through to HAMR writing and finally data storage and read back.

\section{Acknowledgements}

We are grateful to Dr. Mara Strungaru for her aid in the development and testing of the FMR simulation feature within MARS. 

The authors gratefully acknowledge the funding from Transforming Systems through Partnership Programme of the Royal Academy of Engineering under Grant No. TSP1285 and Seagate Technology (Thailand).

J.C and P.C. gratefully acknowledge the financial support from Thailand Science Research and Innovation (TSRI) 

Numerous simulations for this work were undertaken on the Viking Cluster, which is a high performance compute facility provided by the University of York. We are grateful for computational support from the University of York High Performance Computing service, Viking and the Research Computing team.


\begin{thebibliography}{10}
\expandafter\ifx\csname url\endcsname\relax
  \def\url#1{\texttt{#1}}\fi
\expandafter\ifx\csname urlprefix\endcsname\relax\def\urlprefix{URL }\fi
\expandafter\ifx\csname href\endcsname\relax
  \def\href#1#2{#2} \def\path#1{#1}\fi

\bibitem{Richter2007}
H.~J. Richter, The transition from longitudinal to perpendicular recording, J.
  Phys. D: Appl. Phys. 40~(9) (2007) R149--R177.
\newblock \href {https://doi.org/10.1088/0022-3727/40/9/r01}
  {\path{doi:10.1088/0022-3727/40/9/r01}}.

\bibitem{IDEMA}
\url{http://idema.org/?page_id=416}, accessed: 2020-01-29.

\bibitem{SEAGATE}
\url{https://blog.seagate.com/craftsman-ship/hamr-next-leap-forward-now/},
  accessed: 2021-05-11.

\bibitem{Weller2014}
D.~K. Weller, G.~Parker, O.~Mosendz, E.~Champion, B.~Stipe, X.~Wang,
  T.~Klemmer, G.~Ju, A.~Ajan, A {HAMR} media technology roadmap to an areal
  density of 4 {Tb/in$^2$}, IEEE Transactions on Magnetics 50 (2014) 1--8.
\newblock \href {https://doi.org/10.1109/TMAG.2013.2281027}
  {\path{doi:10.1109/TMAG.2013.2281027}}.

\bibitem{Evans2014_field_cooling}
R.~F.~L. Evans, W.~J. Fan, Atomistic simulation of sub-nanosecond
  non-equilibrium field cooling processes for magnetic data storage
  applications, Appl. Phys. Lett. 105~(19) (2014) 192405.
\newblock \href {http://arxiv.org/abs/https://doi.org/10.1063/1.4901959}
  {\path{arXiv:https://doi.org/10.1063/1.4901959}}, \href
  {https://doi.org/10.1063/1.4901959} {\path{doi:10.1063/1.4901959}}.

\bibitem{Vogler2016_AD_optimise}
C.~Vogler, C.~Abert, F.~Bruckner, D.~Suess, D.~Praetorius, Areal density
  optimizations for heat-assisted magnetic recording of high-density media, J.
  Appl. Phys. 119~(22) (2016) 223903.
\newblock \href {http://arxiv.org/abs/https://doi.org/10.1063/1.4953390}
  {\path{arXiv:https://doi.org/10.1063/1.4953390}}, \href
  {https://doi.org/10.1063/1.4953390} {\path{doi:10.1063/1.4953390}}.

\bibitem{Vogler2016_noise}
C.~Vogler, C.~Abert, F.~Bruckner, D.~Suess, D.~Praetorius, Basic noise
  mechanisms of heat-assisted-magnetic recording, J. Appl. Phys. 120~(15)
  (2016) 153901.
\newblock \href {http://arxiv.org/abs/https://doi.org/10.1063/1.4964949}
  {\path{arXiv:https://doi.org/10.1063/1.4964949}}, \href
  {https://doi.org/10.1063/1.4964949} {\path{doi:10.1063/1.4964949}}.

\bibitem{chureemart2017hybrid}
P.~Chureemart, R.~Evans, R.~Chantrell, P.-W. Huang, K.~Wang, G.~Ju,
  J.~Chureemart, Hybrid design for advanced magnetic recording media: Combining
  exchange-coupled composite media with coupled granular continuous media,
  Physical Review Applied 8~(2) (2017) 024016.
\newblock \href {https://doi.org/10.1103/PhysRevApplied.8.024016}
  {\path{doi:10.1103/PhysRevApplied.8.024016}}.

\bibitem{ababei2019anomalous}
R.-V. Ababei, M.~O. Ellis, R.~F. Evans, R.~W. Chantrell, Anomalous damping
  dependence of the switching time in {Fe/FePt} bilayer recording media,
  Physical Review B 99~(2) (2019) 024427.
\newblock \href {https://doi.org/10.1103/PhysRevB.99.024427}
  {\path{doi:10.1103/PhysRevB.99.024427}}.

\bibitem{muthsam2019improving}
O.~Muthsam, F.~Slanovc, C.~Vogler, D.~Suess, Improving the signal-to-noise
  ratio for heat-assisted magnetic recording by optimizing a high/low {Tc}
  bilayer structure, Journal of Applied Physics 126~(12) (2019) 123907.
\newblock \href {https://doi.org/10.1063/1.5119407}
  {\path{doi:10.1063/1.5119407}}.

\bibitem{Ellis2017}
M.~O. Ellis, R.-V. Ababei, R.~Wood, R.~F. Evans, R.~W. Chantrell, Manifestation
  of higher-order inter-granular exchange in magnetic recording media, Applied
  Physics Letters 111~(8) (2017) 082405.
\newblock \href {https://doi.org/10.1063/1.4990604}
  {\path{doi:10.1063/1.4990604}}.

\bibitem{strungaru2020model}
M.~Strungaru, S.~Ruta, R.~F. Evans, R.~W. Chantrell, Model of magnetic damping
  and anisotropy at elevated temperatures: Application to granular fept films,
  Physical Review Applied 14~(1) (2020) 014077.
\newblock \href {https://doi.org/10.1103/PhysRevApplied.14.014077}
  {\path{doi:10.1103/PhysRevApplied.14.014077}}.

\bibitem{meo2020magnetisation}
A.~Meo, W.~Pantasri, W.~Daeng-am, S.~E. Rannala, S.~I. Ruta, R.~W. Chantrell,
  P.~Chureemart, J.~Chureemart, Magnetization dynamics of granular
  heat-assisted magnetic recording media by means of a multiscale model, Phys.
  Rev. B 102 (2020) 174419.
\newblock \href {https://doi.org/10.1103/PhysRevB.102.174419}
  {\path{doi:10.1103/PhysRevB.102.174419}}.

\bibitem{waters2019resolving}
J.~Waters, D.~Kramer, T.~J. Sluckin, O.~Hovorka, Resolving anomalies in the
  critical exponents of {FePt} using finite-size scaling in magnetic fields,
  Physical Review Applied 11~(2) (2019) 024028.
\newblock \href {https://doi.org/10.1103/PhysRevApplied.11.024028}
  {\path{doi:10.1103/PhysRevApplied.11.024028}}.

\bibitem{Richardson2018}
D.~Richardson, S.~Katz, J.~Wang, Y.~K. Takahashi, K.~Srinivasan, A.~Kalitsov,
  K.~Hono, A.~Ajan, M.~Wu, {Near-Tc} ferromagnetic resonance and damping in
  {FePt}-based heat-assisted magnetic recording media, Phys. Rev. Applied 10
  (2018) 054046.
\newblock \href {https://doi.org/10.1103/PhysRevApplied.10.054046}
  {\path{doi:10.1103/PhysRevApplied.10.054046}}.

\bibitem{Tc2014}
S.~Pisana, S.~Jain, J.~W. Reiner, G.~J. Parker, C.~C. Poon, O.~Hellwig, B.~C.
  Stipe, Measurement of the {Curie} temperature distribution in {FePt} granular
  magnetic media, Applied Physics Letters 104~(16) (2014) 162407.
\newblock \href {http://arxiv.org/abs/https://doi.org/10.1063/1.4873543}
  {\path{arXiv:https://doi.org/10.1063/1.4873543}}, \href
  {https://doi.org/10.1063/1.4873543} {\path{doi:10.1063/1.4873543}}.

\bibitem{NEEL}
L.~Néel, Théorie du traînage magnétique des ferromagnétiques en grains
  fins avec applications aux terres cuites, Ann. Geophys. 5 (1949) 99--136.

\bibitem{sLLB-II}
R.~F.~L. Evans, D.~Hinzke, U.~Atxitia, U.~Nowak, R.~W. Chantrell,
  O.~Chubykalo-Fesenko, Stochastic form of the {Landau-Lifshitz-Bloch}
  equation, Phys. Rev. B 85 (2012) 014433.
\newblock \href {https://doi.org/10.1103/PhysRevB.85.014433}
  {\path{doi:10.1103/PhysRevB.85.014433}}.

\bibitem{Hai2014}
H.~Li, J.-G. Zhu, Understanding the impact of {Tc} and {Hk} variation on
  signal-to-noise ratio in heat-assisted magnetic recording, Journal of Applied
  Physics 115~(17) (2014) 17B744.
\newblock \href {http://arxiv.org/abs/https://doi.org/10.1063/1.4868503}
  {\path{arXiv:https://doi.org/10.1063/1.4868503}}, \href
  {https://doi.org/10.1063/1.4868503} {\path{doi:10.1063/1.4868503}}.

\bibitem{Zhu2013}
J.~{Zhu}, H.~{Li}, Understanding signal and noise in heat assisted magnetic
  recording, IEEE Transactions on Magnetics 49~(2) (2013) 765--772.
\newblock \href {https://doi.org/10.1109/TMAG.2012.2231855}
  {\path{doi:10.1109/TMAG.2012.2231855}}.

\bibitem{Eason2014}
K.~{Eason}, H.~T. {Wang}, M.~R. {Elidrissi}, B.~{Xu}, Z.~{Yuan}, K.~S. {Chan},
  Recording performance and comparison of graded-${T_{c}}$ and -${K_{u}}$
  {HAMR} systems, IEEE Transactions on Magnetics 50~(3) (2014) 107--113.
\newblock \href {https://doi.org/10.1109/TMAG.2013.2287509}
  {\path{doi:10.1109/TMAG.2013.2287509}}.

\bibitem{Victoria2013}
R.~H. {Victora}, P.~{Huang}, Simulation of heat-assisted magnetic recording
  using renormalized media cells, IEEE Transactions on Magnetics 49~(2) (2013)
  751--757.
\newblock \href {https://doi.org/10.1109/TMAG.2012.2219300}
  {\path{doi:10.1109/TMAG.2012.2219300}}.

\bibitem{Zhu20012}
H.~Zhu, S.~Thorpe, A.~Windle, The geometrical properties of irregular
  two-dimensional {Voronoi} tessellations, Philosophical Magazine A: Physics of
  Condensed Matter, Structure, Defects and Mechanical Properties 81~(12) (2001)
  2765--2783, cited By 98.
\newblock \href {https://doi.org/10.1080/01418610010032364}
  {\path{doi:10.1080/01418610010032364}}.

\bibitem{GRENESTEDT1998}
J.~L. Grenestedt, K.~Tanaka, Influence of cell shape variations on elastic
  stiffness of closed cell cellular solids, Scripta Materialia 40~(1) (1998) 71
  -- 77.
\newblock \href {https://doi.org/https://doi.org/10.1016/S1359-6462(98)00401-1}
  {\path{doi:https://doi.org/10.1016/S1359-6462(98)00401-1}}.

\bibitem{Lloyd}
S.~{Lloyd}, Least squares quantization in {PCM}, IEEE Transactions on
  Information Theory 28~(2) (1982) 129--137.
\newblock \href {https://doi.org/10.1109/TIT.1982.1056489}
  {\path{doi:10.1109/TIT.1982.1056489}}.

\bibitem{Liu2009}
Y.~Liu, W.~Wang, B.~L\'{e}vy, F.~Sun, D.-M. Yan, L.~Lu, C.~Yang, On centroidal
  {Voronoi} tessellation—energy smoothness and fast computation, ACM Trans.
  Graph. 28~(4) (Sep. 2009).
\newblock \href {https://doi.org/10.1145/1559755.1559758}
  {\path{doi:10.1145/1559755.1559758}}.

\bibitem{FAZEKAS2002}
A.~Fazekas, R.~Dendievel, L.~Salvo, Y.~Bréchet, Effect of microstructural
  topology upon the stiffness and strength of {2D} cellular structures,
  International Journal of Mechanical Sciences 44~(10) (2002) 2047 -- 2066.
\newblock \href {https://doi.org/https://doi.org/10.1016/S0020-7403(02)00171-6}
  {\path{doi:https://doi.org/10.1016/S0020-7403(02)00171-6}}.

\bibitem{Garvois}
A.~Gervois, L.~Oger, P.~Richard, J.~P. Troadec, Voronoi and radical
  tessellations of packings of spheres, in: P.~M.~A. Sloot, A.~G. Hoekstra,
  C.~J.~K. Tan, J.~J. Dongarra (Eds.), Computational Science --- ICCS 2002,
  Springer Berlin Heidelberg, Berlin, Heidelberg, 2002, pp. 95--104.
\newblock \href {https://doi.org/10.1007/3-540-47789-6_10}
  {\path{doi:10.1007/3-540-47789-6_10}}.

\bibitem{Su-Peng2009}
S.-P. Yu, Y.-H. Liu, A.-C. Sun, J.-H. Hsu, Determining the size distribution of
  magnetic nanoparticles based on analysis of magnetization curves, Journal of
  Applied Physics 106~(10) (2009) 103905.
\newblock \href {http://arxiv.org/abs/https://doi.org/10.1063/1.3259424}
  {\path{arXiv:https://doi.org/10.1063/1.3259424}}, \href
  {https://doi.org/10.1063/1.3259424} {\path{doi:10.1063/1.3259424}}.

\bibitem{Ganping2006}
G.~Ju, H.~Zhou, R.~Chantrell, B.~Lu, D.~Weller, Direct probe of anisotropy
  field dispersion in perpendicular media, Journal of Applied Physics 99~(8)
  (2006) 083902.
\newblock \href {http://arxiv.org/abs/https://doi.org/10.1063/1.2189025}
  {\path{arXiv:https://doi.org/10.1063/1.2189025}}, \href
  {https://doi.org/10.1063/1.2189025} {\path{doi:10.1063/1.2189025}}.

\bibitem{ALSAYEDNOOR2016}
J.~Alsayednoor, P.~Harrison, Evaluating the performance of microstructure
  generation algorithms for {2d} foam-like representative volume elements,
  Mechanics of Materials 98 (2016) 44 -- 58.
\newblock \href {https://doi.org/https://doi.org/10.1016/j.mechmat.2016.04.001}
  {\path{doi:https://doi.org/10.1016/j.mechmat.2016.04.001}}.

\bibitem{VORO}
C.~H. Rycroft, Voro++: A three-dimensional {Voronoi} cell library in {C++},
  Chaos 19~(4) (2009) 041111.
\newblock \href {http://arxiv.org/abs/https://doi.org/10.1063/1.3215722}
  {\path{arXiv:https://doi.org/10.1063/1.3215722}}, \href
  {https://doi.org/10.1063/1.3215722} {\path{doi:10.1063/1.3215722}}.

\bibitem{Peng2011}
Y.~Peng, X.~W. Wu, J.~Pressesky, G.~P. Ju, W.~Scholz, R.~Chantrell, Cluster
  size and exchange dispersion in perpendicular magnetic media, Journal of
  Applied Physics 109~(12) (2011) 123907.
\newblock \href {http://arxiv.org/abs/https://doi.org/10.1063/1.3596806}
  {\path{arXiv:https://doi.org/10.1063/1.3596806}}, \href
  {https://doi.org/10.1063/1.3596806} {\path{doi:10.1063/1.3596806}}.

\bibitem{Sokalski2009}
V.~Sokalski, D.~E. Laughlin, J.-G. Zhu, Experimental modeling of intergranular
  exchange coupling for perpendicular thin film media, Applied Physics Letters
  95~(10) (2009) 102507.
\newblock \href {https://doi.org/10.1063/1.3226638}
  {\path{doi:10.1063/1.3226638}}.

\bibitem{Sergiu_th}
S.~Ruta, Study of interaction effects in magnetic granular systems for
  recording media application, Ph.D. thesis, University of York (9 2017).

\bibitem{Callen1966}
H.~Callen, E.~Callen, The present status of the temperature dependence of
  magnetocrystalline anisotropy, and the {l(l+1)2} power law, J. Phys. Chem.
  Solids 27~(8) (1966) 1271 -- 1285.
\newblock \href {https://doi.org/https://doi.org/10.1016/0022-3697(66)90012-6}
  {\path{doi:https://doi.org/10.1016/0022-3697(66)90012-6}}.

\bibitem{Mryasov2005}
O.~Mryasov, U.~Nowak, K.~Guslienko, R.~Chantrell, Temperature-dependent
  magnetic properties of {FePt}: Effective spin hamiltonian model, Europhysics
  Letters 69~(5) (2005) 805--811.
\newblock \href {https://doi.org/10.1209/epl/i2004-10404-2}
  {\path{doi:10.1209/epl/i2004-10404-2}}.

\bibitem{Saslow2009}
W.~M. Saslow, {Landau–Lifshitz} or {Gilbert} damping? {That} is the question,
  Journal of Applied Physics 105~(7) (2009) 07D315.
\newblock \href {http://arxiv.org/abs/https://doi.org/10.1063/1.3077204}
  {\path{arXiv:https://doi.org/10.1063/1.3077204}}, \href
  {https://doi.org/10.1063/1.3077204} {\path{doi:10.1063/1.3077204}}.

\bibitem{Garanin1990}
D.~A. Garanin, V.~V. Ishchenko, L.~V. Panina, Dynamics of an ensemble of
  single-domain magnetic particles, Theoretical and Mathematical Physics 82~(2)
  (1990) 169--179.
\newblock \href {https://doi.org/10.1007/BF01079045}
  {\path{doi:10.1007/BF01079045}}.

\bibitem{Garanin1997}
D.~A. Garanin, {Fokker-Planck} and {Landau-Lifshitz-Bloch} equations for
  classical ferromagnets, Phys. Rev. B 55 (1997) 3050--3057.
\newblock \href {https://doi.org/10.1103/PhysRevB.55.3050}
  {\path{doi:10.1103/PhysRevB.55.3050}}.

\bibitem{Oksana_free_energy}
O.~Chubykalo-Fesenko, U.~Nowak, R.~W. Chantrell, D.~Garanin, Dynamic approach
  for micromagnetics close to the {Curie} temperature, Phys. Rev. B 74 (2006)
  094436.
\newblock \href {https://doi.org/10.1103/PhysRevB.74.094436}
  {\path{doi:10.1103/PhysRevB.74.094436}}.

\bibitem{Evans2014}
R.~F.~L. Evans, W.~J. Fan, P.~Chureemart, T.~A. Ostler, M.~O.~A. Ellis, R.~W.
  Chantrell, Atomistic spin model simulations of magnetic nanomaterials, J.
  Phys-Condens Mat. 26~(10) (2014) 103202.
\newblock \href {https://doi.org/10.1088/0953-8984/26/10/103202}
  {\path{doi:10.1088/0953-8984/26/10/103202}}.

\bibitem{Matt-LLB}
M.~Ellis, Simulations of magnetic reversal in granular recording media, Ph.D.
  thesis, University of York (9 2015).

\bibitem{Vogler2014_LLB}
C.~Vogler, C.~Abert, F.~Bruckner, D.~Suess, {Landau-Lifshitz-Bloch} equation
  for exchange-coupled grains, Phys. Rev. B 90 (2014) 214431.
\newblock \href {https://doi.org/10.1103/PhysRevB.90.214431}
  {\path{doi:10.1103/PhysRevB.90.214431}}.

\bibitem{Chantrell2001a}
R.~Chantrell, N.~Walmsley, J.~Gore, M.~Maylin, Calculations of the
  susceptibility of interacting superparamagnetic particles, Physical Review B
  63~(2) (2001) 024410.
\newblock \href {https://doi.org/10.1103/PhysRevB.63.024410}
  {\path{doi:10.1103/PhysRevB.63.024410}}.

\bibitem{Pfeiffer1990}
H.~Pfeiffer, Determination of anisotropy field distribution in particle
  assemblies taking into account thermal fluctuations, physica status solidi
  (a) 118~(1) (1990) 295--306.
\newblock \href
  {http://arxiv.org/abs/https://onlinelibrary.wiley.com/doi/pdf/10.1002/pssa.2211180133}
  {\path{arXiv:https://onlinelibrary.wiley.com/doi/pdf/10.1002/pssa.2211180133}},
  \href {https://doi.org/https://doi.org/10.1002/pssa.2211180133}
  {\path{doi:https://doi.org/10.1002/pssa.2211180133}}.

\bibitem{Stoner-Wohlfarth}
E.~C. Stoner, E.~P. Wohlfarth, A mechanism of magnetic hysteresis in
  heterogeneous alloys, Phil. Trans. R. Soc. A 240~(826) (1948) 599--642.
\newblock \href {https://doi.org/10.1098/rsta.1948.0007}
  {\path{doi:10.1098/rsta.1948.0007}}.

\bibitem{Li2013}
H.~{Li}, J.~{Zhu}, The role of media property distribution in {HAMR SNR}, IEEE
  Transactions on Magnetics 49~(7) (2013) 3568--3571.
\newblock \href {https://doi.org/10.1109/TMAG.2012.2234086}
  {\path{doi:10.1109/TMAG.2012.2234086}}.

\bibitem{Zhu2014}
J.-G.~J. Zhu, H.~Li, Signal-to-noise ratio impact of grain-to-grain heating
  variation in heat assisted magnetic recording, Journal of Applied Physics
  115~(17) (2014) 17B747.
\newblock \href {http://arxiv.org/abs/https://doi.org/10.1063/1.4867607}
  {\path{arXiv:https://doi.org/10.1063/1.4867607}}, \href
  {https://doi.org/10.1063/1.4867607} {\path{doi:10.1063/1.4867607}}.

\bibitem{Rong2006}
C.-b. Rong, D.~Li, V.~Nandwana, N.~Poudyal, Y.~Ding, Z.~Wang, H.~Zeng, J.~Liu,
  Size-dependent chemical and magnetic ordering in {L10-FePt} nanoparticles,
  Advanced Materials 18~(22) (2006) 2984--2988.
\newblock \href
  {http://arxiv.org/abs/https://onlinelibrary.wiley.com/doi/pdf/10.1002/adma.200601904}
  {\path{arXiv:https://onlinelibrary.wiley.com/doi/pdf/10.1002/adma.200601904}},
  \href {https://doi.org/https://doi.org/10.1002/adma.200601904}
  {\path{doi:https://doi.org/10.1002/adma.200601904}}.

\bibitem{Hovorka2012a}
O.~Hovorka, S.~Devos, Q.~Coopman, W.~J. Fan, C.~J. Aas, R.~F.~L. Evans,
  X.~Chen, G.~Ju, R.~W. Chantrell, The {Curie} temperature distribution of
  {FePt} granular magnetic recording media, Applied Physics Letters 101~(5)
  (2012) 052406.
\newblock \href {https://doi.org/10.1063/1.4740075}
  {\path{doi:10.1063/1.4740075}}.

\bibitem{Hernandez2016}
S.~Hernandez, P.~Krivosik, P.~W. Huang, W.~R. Eppler, T.~Rausch, E.~Gage,
  Parametric comparison of modelled and measured heat-assisted magnetic
  recording using a common signal-to-noise metric, IEEE Transactions on
  Magnetics 52~(7) (2016) 1--4.
\newblock \href {https://doi.org/10.1109/TMAG.2016.2528224}
  {\path{doi:10.1109/TMAG.2016.2528224}}.

\bibitem{Hernandez2017}
S.~Hernandez, P.-L. Lu, S.~Granz, P.~Krivosik, P.-W. Huang, W.~Eppler,
  T.~Rausch, E.~Gage, Using ensemble waveform analysis to compare heat assisted
  magnetic recording characteristics of modeled and measured signals, IEEE
  Transactions on Magnetics 53~(2) (2017) 1--6.
\newblock \href {https://doi.org/10.1109/TMAG.2016.2612230}
  {\path{doi:10.1109/TMAG.2016.2612230}}.

\bibitem{hashmi2015reference}
S.~Hashmi, {Reference Module in Materials Science and Materials Engineering},
  Elsevier, Amsterdam, 2015.

\bibitem{yalcin2013ferromagnetic}
O.~Yalcin, {Ferromagnetic Resonance: Theory and Applications}, Intech, Croatia,
  2013.

\bibitem{eriksson2017atomistic}
O.~Eriksson, {Atomistic Spin Dynamics: Foundations and Applications}, Oxford
  University Press, Oxford, United Kingdom, 2017.

\bibitem{kittel}
C.~Kittel, {Introduction To Solid State Physics}, Wiley, Hoboken, NJ, 2005.

\end{thebibliography}

\end{document}


\date{3 September 2021}
\maketitle
\newpage

\section{Supplementary Note 1 - General solver testing}

\subsection{Verification of the Heun scheme}

The dynamic solvers utilise the Heun integration scheme due to its ability to provide the required Stratonovich solution of stochastic equations. However, first this method must be shown to accurately solve the deterministic equations of motion. For this test analytical solutions were derived for specific cases for both the LLG and LLB solvers. For this test a single grain with an initial magnetisation along the x-axis under the influence of an applied field directed along the z-axis \cite{Hannay} is used, the time evolution for the LLG solver is then given by:

\begin{equation}
    \begin{aligned}
        M_x(t) &= \sech\bigg(\frac{\gamma \alpha H}{1+\alpha^2}t\bigg) \cos\bigg(\frac{\gamma H}{1+\alpha^2}t\bigg) \\
        M_x(t) &= \sech\bigg(\frac{\gamma \alpha H}{1+\alpha^2}t\bigg) \sin\bigg(\frac{\gamma H}{1+\alpha^2}t\bigg) \\
        M_z(t) &= \tanh\bigg(\frac{\gamma \alpha H}{1+\alpha^2}t\bigg) \; .
    \end{aligned}
    \label{eq:LLG_analyt}
\end{equation}

For the LLB solver the time evolution is:
\begin{equation}
    \begin{aligned}
        M_x(t) &= \sech\bigg(\frac{\gamma \alpha_{\perp} Ht}{m^2}\bigg) \cos(\gamma \alpha_{\perp} Ht) \\
        M_y(t) &= \sech\bigg(\frac{\gamma \alpha_{\perp} Ht}{m^2}\bigg) \sin(\gamma \alpha_{\perp} Ht) \\
        M_z(t) &= \tanh\bigg(\frac{\gamma \alpha_{\perp} Ht}{m^2}\bigg) \; .
    \end{aligned}
    \label{eq:LLB_analyt}
\end{equation}

The time evolution for the LLG and LLB along with the obtained errors are shown in supplementary figure \ref{fig:Analytics}. The maximum error obtained for the LLG is around $10^{-2}$, while for the LLB it is around $10^{-5}$. This is to be expected as the timestep utilised in the LLB is 1 fs and 1 ps in the LLG. The form of these errors is characteristic of a correctly implemented Heun integration scheme.

\begin{figure}[htb]
    \centering
    \includegraphics[width=0.6\textwidth]{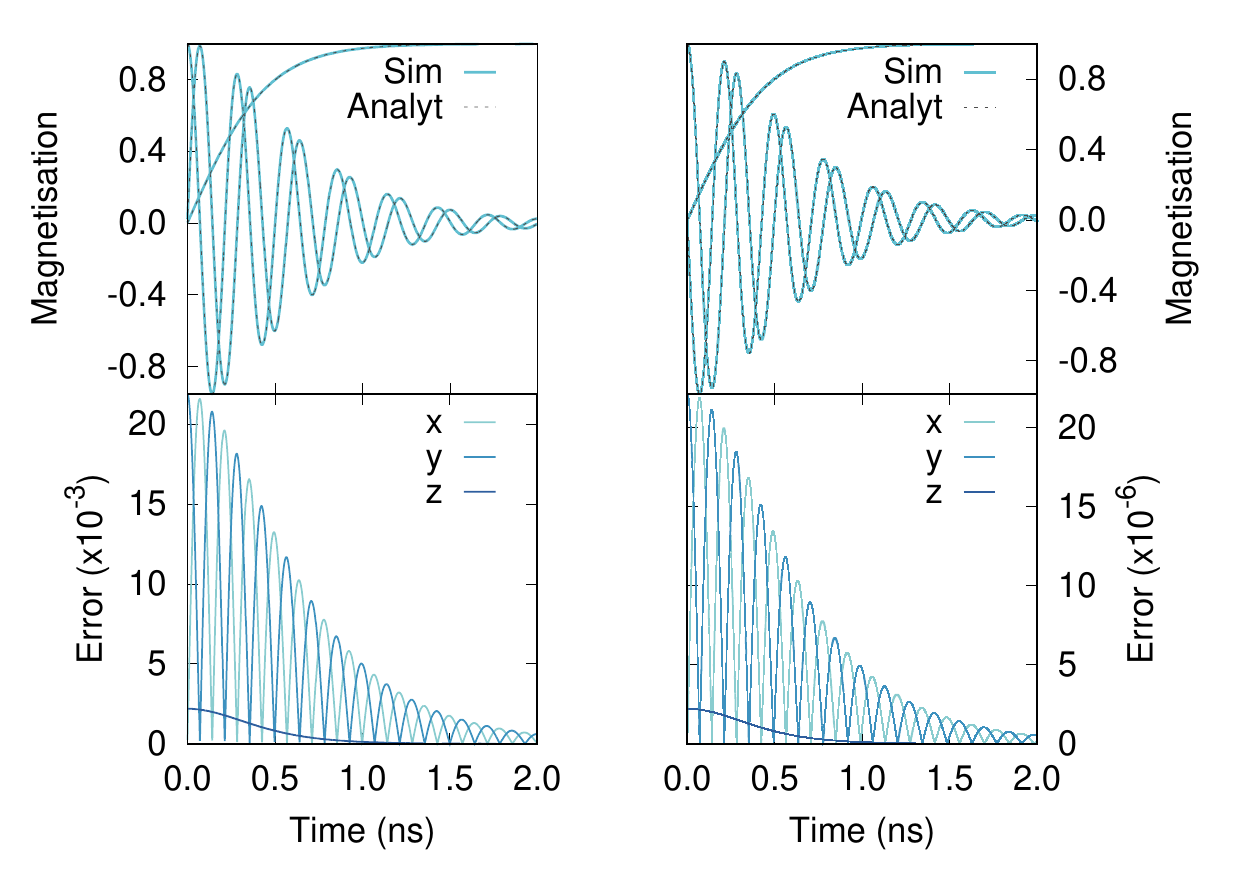}
    \caption{Comparison of the numerical simulation and analytical solution for the LLG (left) and LLB (right) solvers. Both systems are in the presence of a $400\pi\ Oe$ field and the timesteps of integration used are $\Delta =$ \SI{1}{ps} for the LLG and $\Delta t=$ \SI{1}{fs} for the LLB.}
    \label{fig:Analytics}
\end{figure}

\subsection{Angular dependence of the coercivity} \label{sec:Angular dependence of the coercivity}

Verification of the implementation of the uniaxial anisotropy for all solvers is performed via a simple test which makes use of the Stoner-Wohlfarth model. Here, the Stoner-Wohlfarth particle describes the behaviour of a single grain at zero Kelvin under the influence of an applied field. The angular dependence of the coercivity is very well known \cite{Stoner-Wohlfarth} and thus makes a very useful property for comparison and verification of a code. This test verifies the deterministic behaviour of the LLG and LLB by ensuring the easy axis profile provides a coercivity of $H_k = \frac{2K}{M_s}$ as is known analytically. 

The grain is initially magnetised along the z-axis and the applied field strength is varied from $H=1.5H_k$ to $H=-1.5H_k$ and back, in steps of $0.005H_k$. This process is repeated for a range of field directions ranging from $90^\circ$ to $0^\circ$. The projection of the magnetisation on the field direction is then plotted as a function of the applied field strength as shown in supplementary figure \ref{fig:Hysteresis} for the LLG, LLB and kMC solvers. As expected the same profiles are obtained irrespective of which of the three solvers are used and these profiles also agree exactly with those of Stoner and Wohlfarth's solution.

\begin{figure}[htb]
    \centering
    \includegraphics[width=0.6\textwidth]{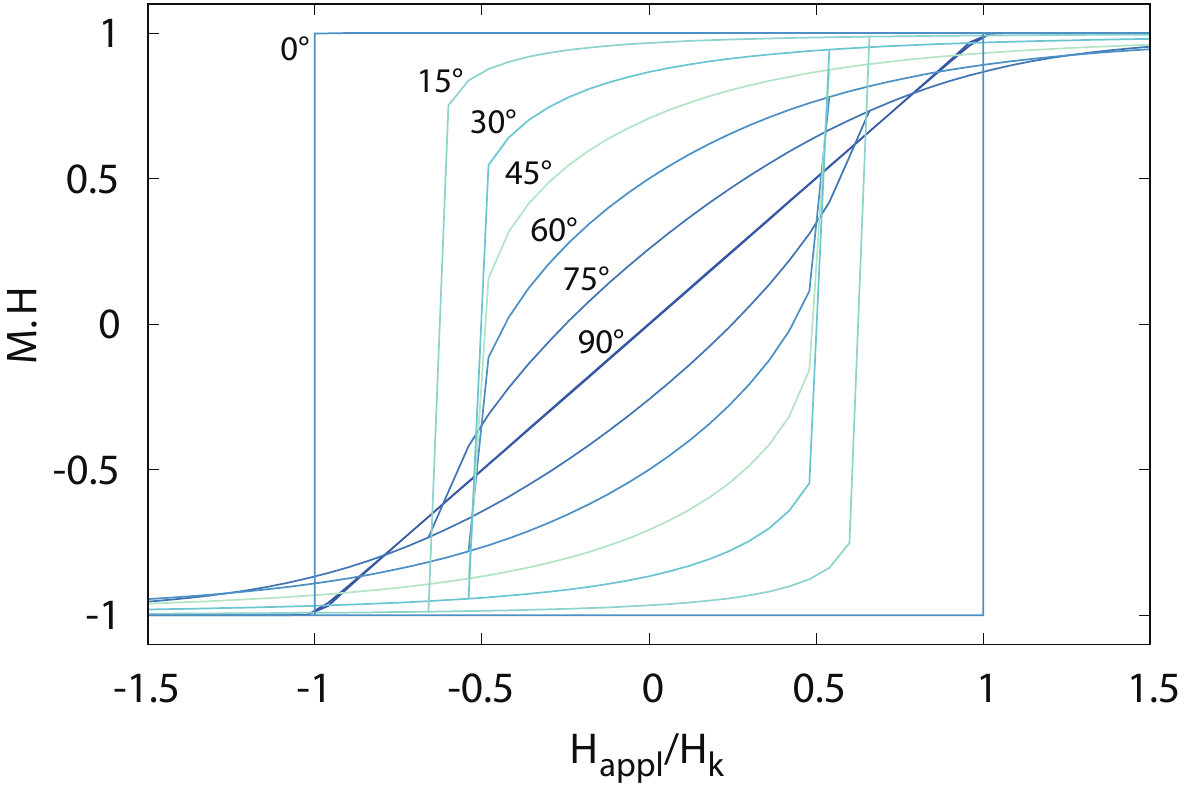}
    \caption{Alignment magnetisation for a single grain under an applied field for various angles from the easy axis. The $0^\circ$ and $90^\circ$ profiles are simulated with a small deviation from the labelled value. This is done as  at perfect alignment, in the absence of thermal noise, there is zero torque which prevents any change in the magnetisation alignment.}
    \label{fig:Hysteresis}
\end{figure}

\subsection{Boltzmann distributions for an ensemble of non-interacting grains}

To verify the stochastic behaviour of the sLLG and sLLB-II solvers a test which utilises thermal effects is required. The most simple test available is the reproduction of the Boltzmann distribution for an ensemble of non-interacting grains, with uniaxial anisotropy in the absence of an applied field. The Boltzmann distribution for magnetisation as a function of polar angle for the LLG is simple to obtain and is given by:

\begin{equation}
    P(\theta _M) = \frac{\sin(\theta_m)\exp\bigg(-\frac{KV\sin(\theta_M)^2}{k_B T}\bigg)}{\int_{0}^{\pi/2} \sin(\theta_m)\exp\bigg(-\frac{KV\sin(\theta_M)^2}{k_B T}\bigg)d\theta_M} \; ,
\end{equation}

where $\theta_M$ is the angle of magnetisation with respect to the easy axis. The ensemble is allowed to evolve over time for $10^6$ steps and the angle of magnetisation is recorded at each step. The results obtained by the sLLG and the analytical solution are shown in supplementary figure \ref{fig:LLG_Boltzmann} for an initial magnetisation direction parallel to the easy axis direction. There is excellent agreement between these results, showing correct implementation of thermal effects in the sLLG.

\begin{figure}
    \centering
    \includegraphics[width=0.6\textwidth]{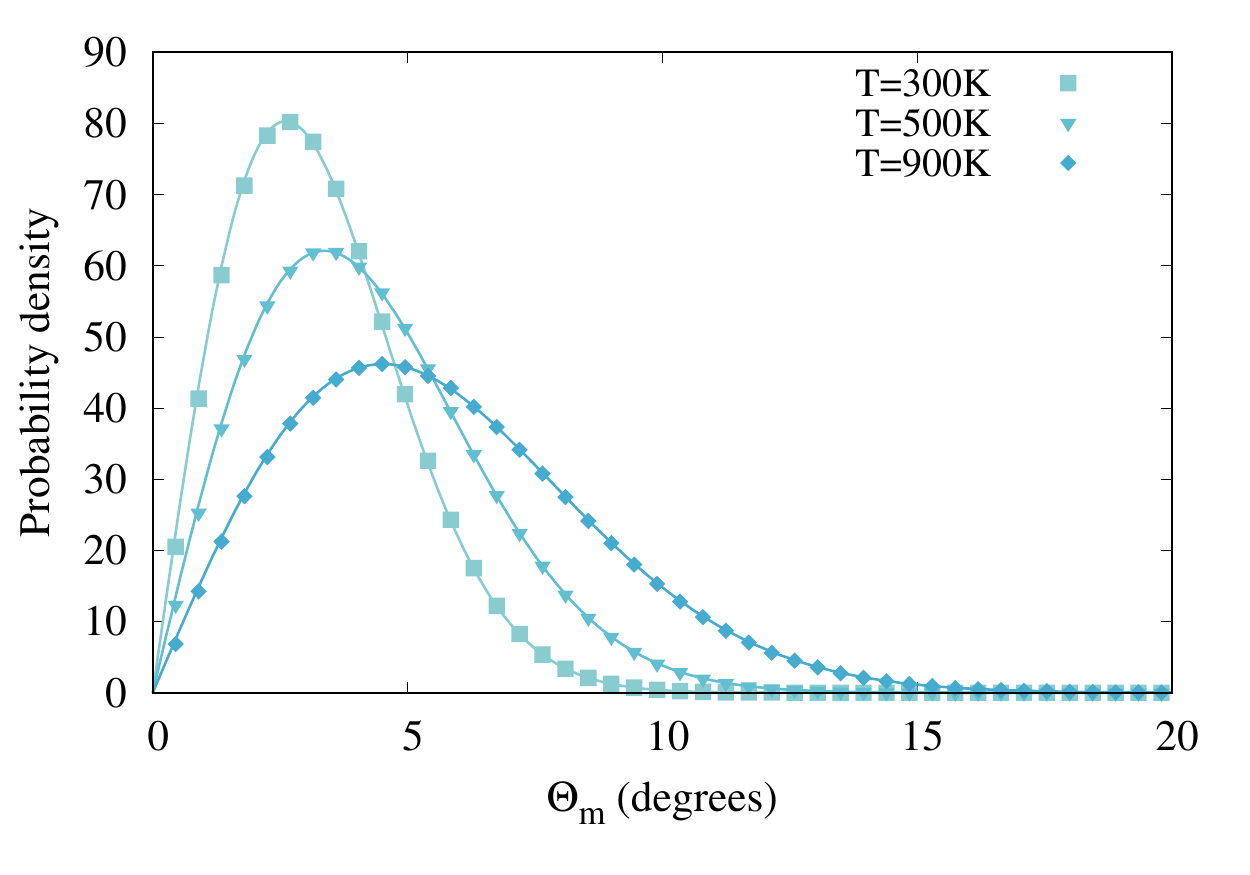}
    \caption{Angular probability distributions for a non-interacting ensemble of grains with uniaxial anisotropy for 300K, 500K and 900K. The analytical Boltzmann distributions are represented by the solid lines, the simulated data is represented by points.}
    \label{fig:LLG_Boltzmann}
\end{figure}

For the LLB equation the free energy of the system is defined as \cite{Kazantseva2008_FePt,Oksana_free_energy}:

\begin{equation}
    \frac{F}{M^0_SV} =
        \left \{
    \begin{aligned}
        \frac{m^2_x+m^2_y}{2\Chiperp} + \frac{(m^2-m^2_e)^2}{8\Chipara m^2_e}, 
 && \text{if } T \leq T_c \\
        \frac{m^2_x+m^2_y}{2\Chiperp} + \frac{3}{20\Chipara} \frac{T_c}{T-T_c} \bigg(m^2 +\frac{5}{3}\frac{T-T_c}{T_c}\bigg)^2, 
 && \text{otherwise.} \\
    \end{aligned} \right.
\end{equation}

Where the first term provides uniaxial anisotropy and the second controls the magnetisation length. The expected Boltzmann distribution is of the form:

\begin{equation}
    P(|m|) \propto m^2 \exp{\bigg(-\frac{F}{k_BT}\bigg)} \; .
\end{equation}

Probability distributions along $m_z$ and $m_x$ for different temperatures along with the corresponding analytical solutions are plotted in supplementary figure \ref{fig:LLB_boltzmann}, all results show strong agreement with the analytical solutions. The simulated system has a Curie point of \SI{685.14}{K}, includes anisotropy with an initial random uniformly distributed magnetisation and has been evolved over time for \SI{1}{ns} ($10^{6}$ steps). The results show the ability of the sLLB-II solver to describe the loss of ferromagnetic behaviour as the temperature approaches and exceeds the Curie point, indicated by the formation of a single peak once the temperature exceeds the Curie point.

\begin{figure}[H]
    \centering
    \includegraphics[width=0.6\linewidth]{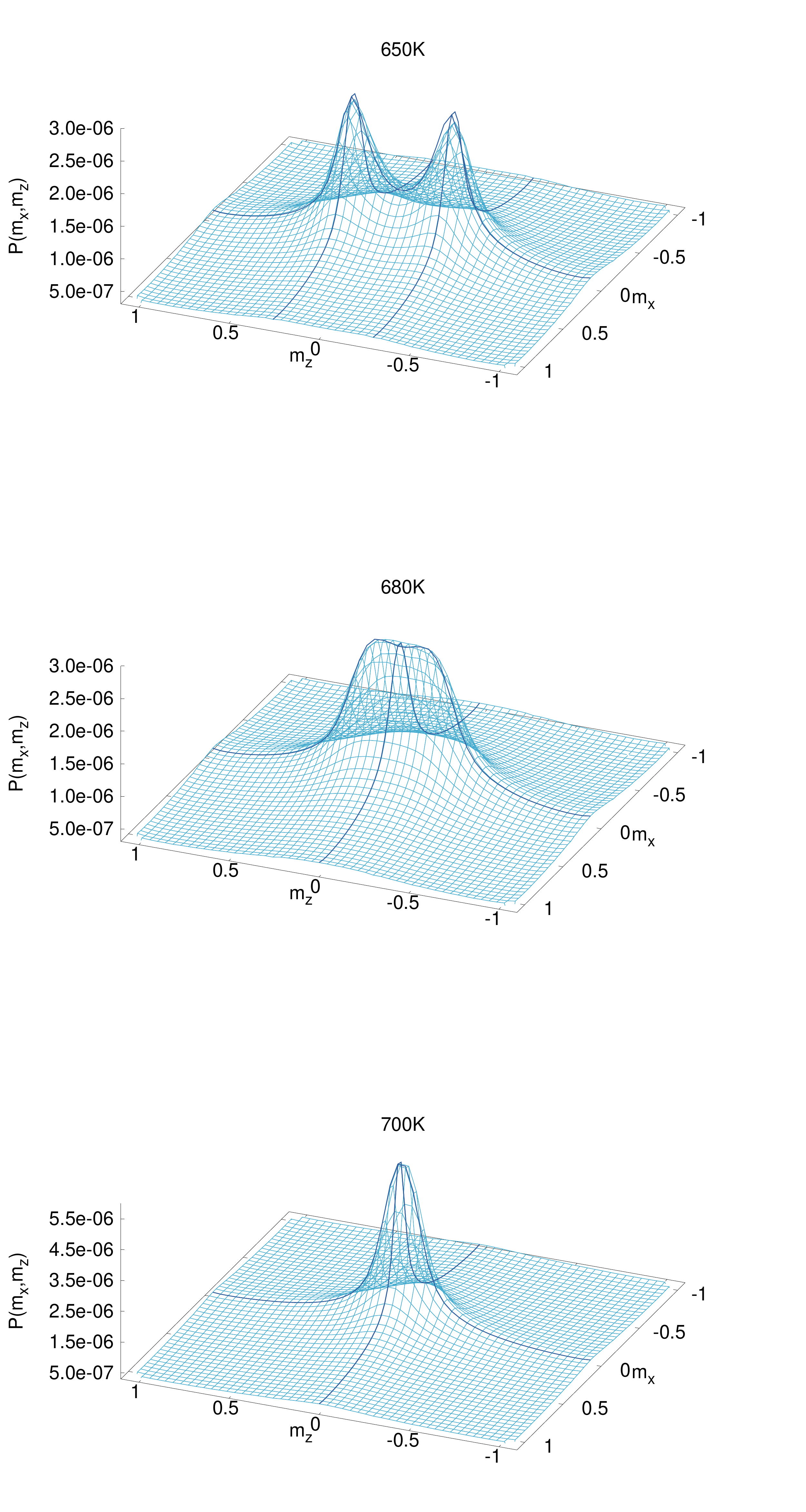}
    \caption{Anisotropic distributions of magnetisation as functions of $m_x$ and $m_z$ (light blue grid) compared to analytical solutions for slices of $m_y$ (dark blue lines).}
    \label{fig:LLB_boltzmann}
\end{figure}

\newpage

\section{Supplementary Note 2 - Landau-Lifshitz-Bloch solver specific testing}

\subsection{Determination of the Curie temperature}

\begin{figure}[htb]
    \centering
    \captionsetup{justification=centering}
    \includegraphics[width=0.6\textwidth]{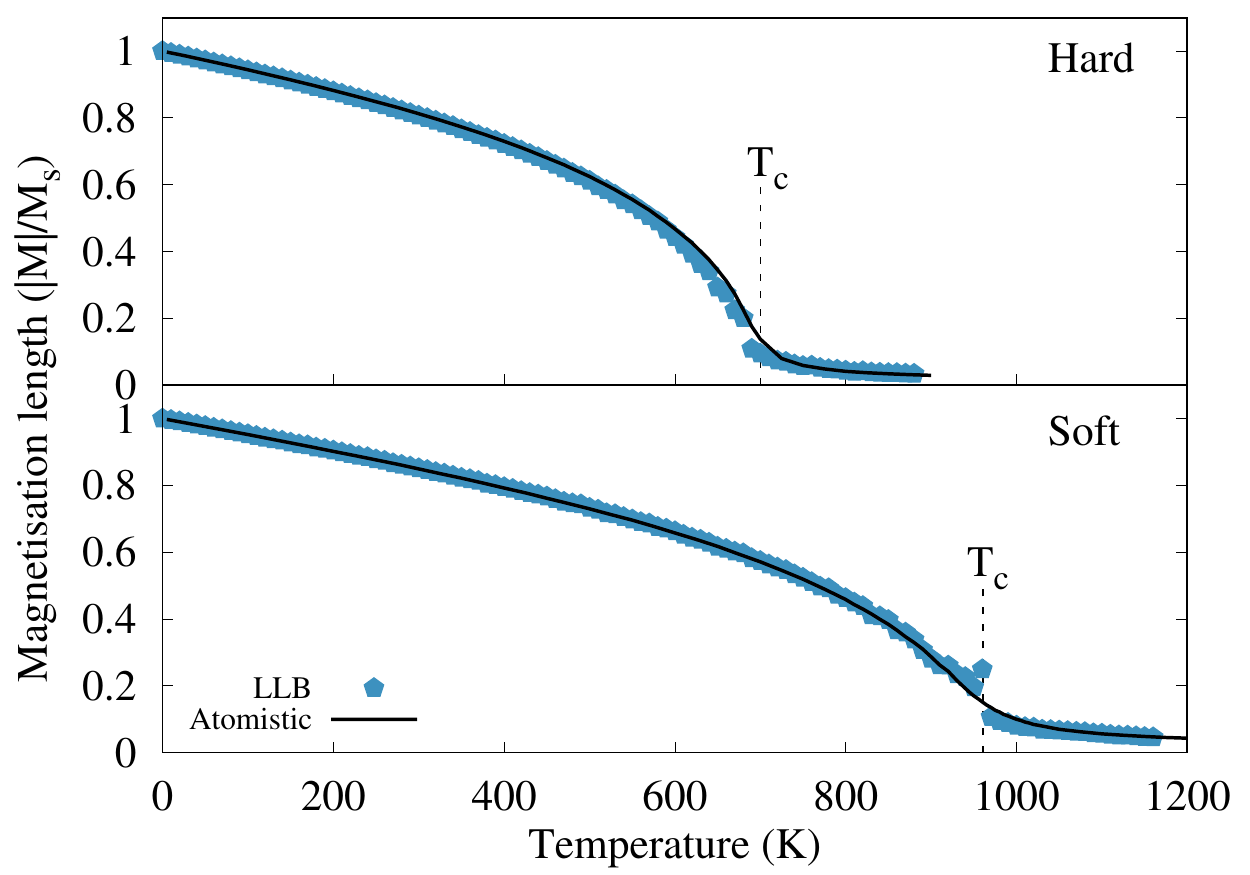}
    \caption{Magnetisation as a function of temperature for the hard and soft materials performed using LLB solver.}
    \label{fig:LLB_mVt}
\end{figure}

The LLG conserves magnetisation length assuming unity for all temperatures, thus the model starts to break down at temperatures approaching $T_c$. The LLB allows for variable magnetisation lengths enabling it to simulate systems at and beyond their Curie points. This test was devised to verify the treatment of the magnetisation length by the LLB. The temperature dependence of the magnetisation is determined via parameterisation obtained using atomistic simulations. Two systems are simulated, one with a hard magnetic material the other a soft material. The hard material is modelled via the perpendicular susceptibility while the soft material utilises the Callen-Callen method. The systems are heated from \SI{0}{K} to above the highest Curie point of the materials. At each temperature the system is simulated until an equilibrium is reached at which point the magnetisation is recorded and the temperature increased.
supplementary figure \ref{fig:LLB_mVt} shows the results obtained for the hard and soft materials via LLB simulation compared to those obtained by atomistic simulation. There is excellent agreement between the results, showing that MARS is capable of reproducing the correct behaviour of the magnetisation length.

\subsection{Longitudinal relaxation}

\begin{figure}[htb]
    \centering
    \captionsetup{justification=centering}
    \includegraphics[width=0.6\textwidth]{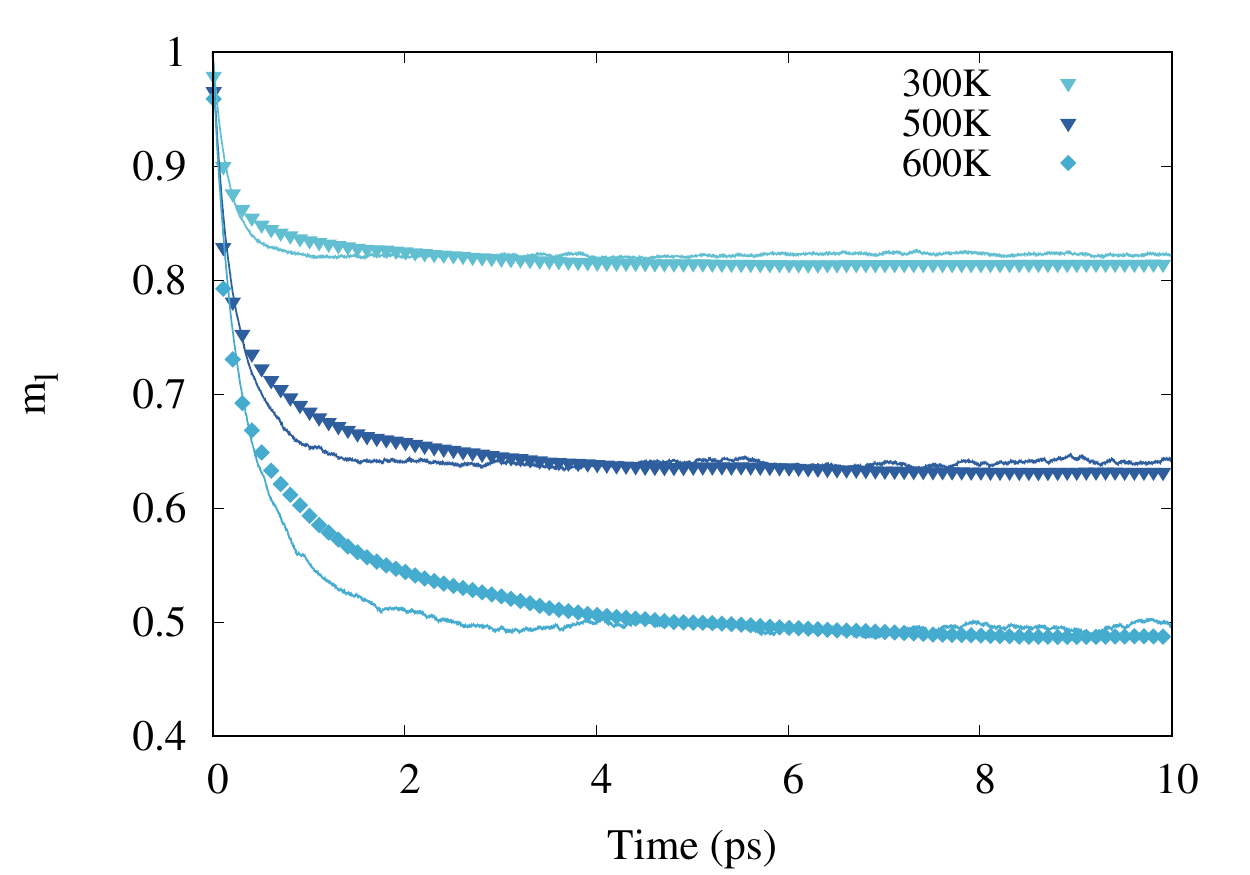}
    \caption{Comparison of micromagnetic (lines) and atomistic (dots) results of the magnetisation length as a function of time, for a system undergoing thermal relaxation. The Curie temperature of the simulated system is 680K.}
    \label{fig:Transverse}
\end{figure}

A system of non-interacting identical grains is initially aligned at $30^{\circ}$ from the z-axis, placing it in a non-equilibrium state. The system is then allowed to thermally relax, in the absence of an applied field, for a range of system temperatures and the magnetisation length is recorded at each step. supplementary figure \ref{fig:Transverse} shows the comparison of the results obtained via MARS and atomistic simulations. The results show excellent agreement. As expected there is a small discrepancy for very short timescales where the rate of relaxation is greatest \cite{Kazantseva2008_FePt}, however, both results show identical equilibrium lengths within only \SI{5}{ps}. 

\section{Supplementary Note 3 - Kinetic Monte Carlo solver specific testing}

\subsection{Coercivity as a function of sweep rate}

\begin{figure}[H]
    \centering
    \includegraphics[width=0.6\textwidth]{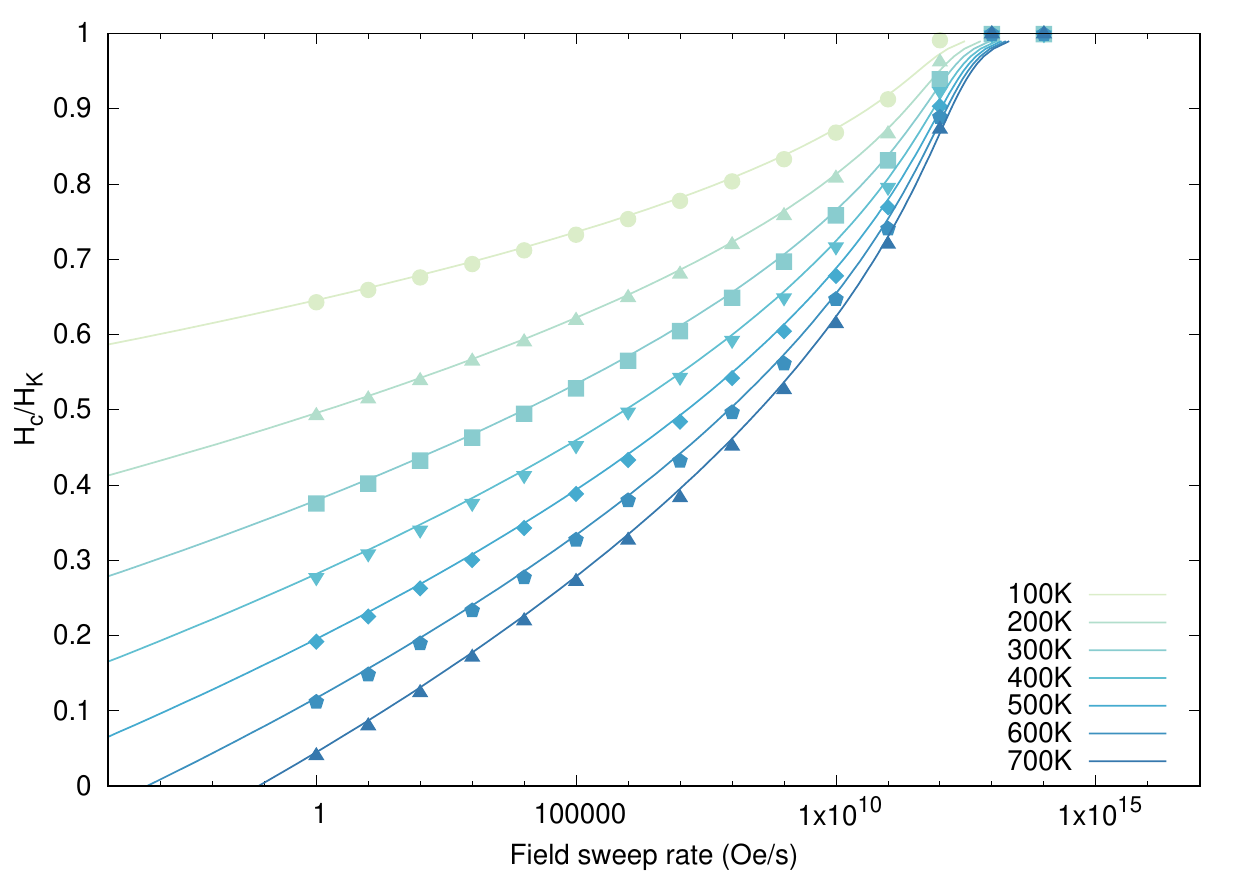}
    \caption{Coercivity as function of applied field sweep rate for the easy axis aligned with the field for various temperatures $100K$ to $700K$. The simulation results (dots) show good agreement with the theoretical prediction (lines).}
    \label{fig:KMC_test}
\end{figure}

In the presence of thermal effects the coercivity depends on the applied field sweep rate. The coercivity as a function of the sweep rate was determined empirically by Sharrock \cite{Sweep_rate_Flanders} and then derived theoretically by Chantrell \cite{Sweep_rate_Roy} under the assumption of constant attempt frequency and an easy axis parallel to the applied field. A simple test for the kMC solver consists of the simulation of hysteresis profiles for a range of sweep rates. Supplementary figure \ref{fig:KMC_test} shows the comparison between the simulated results and the theoretical prediction. The results agree strongly with the theory verifying the implementation of the kMC for thermal systems.

\bibliographystyle{elsarticle-num}